%% file: main.tex
\documentclass[sigconf]{acmart}
\usepackage{multirow}
\usepackage{makecell}
\usepackage{soul}
\usepackage{cleveref}
\usepackage{colortbl}
\usepackage{subcaption}
\usepackage{siunitx}
\usepackage{enumitem}
\usepackage{pifont}
\usepackage{xspace}

\AtBeginDocument{%
  }

\newcommand{\mainfinding}[2]{%
  \item[\hspace{1em}\ding{\numexpr201+#1\relax}] #2%
}

\newcommand{\noai}{\textsc{Control}\xspace}
\newcommand{\singleai}{\textsc{AI-Single}\xspace}
\newcommand{\multipleai}{\textsc{AI-Multiple}\xspace}

\newcommand{\revise}[1]{\textcolor{black}{#1}}

\copyrightyear{2026}
\acmYear{2026}
\setcopyright{cc}
\setcctype{by}
\acmConference[CHI '26]{Proceedings of the 2026 CHI Conference on Human Factors in Computing Systems}{April 13--17, 2026}{Barcelona, Spain}
\acmBooktitle{Proceedings of the 2026 CHI Conference on Human Factors in Computing Systems (CHI '26), April 13--17, 2026, Barcelona, Spain}
\acmPrice{}
\acmDOI{10.1145/3772318.3791727}
\acmISBN{979-8-4007-2278-3/2026/04}

\begin{document}

\title[A Randomized Experiment Investigating the Effects of AI on Close Reading]{What Does AI Do for Cultural Interpretation? \\ A Randomized Experiment on Close Reading Poems with Exposure to AI Interpretation}


\author{Jiayin Zhi}
\orcid{0009-0006-9290-7356}
\email{jzhi@uchicago.edu}
\affiliation{%
 \institution{University of Chicago}
 \city{Chicago}
 \state{Illinois}
 \country{USA}}

\author{Hoyt Long}
\orcid{0000-0002-8562-5426}
\email{hoytlong@uchicago.edu}
\affiliation{%
 \institution{University of Chicago}
 \city{Chicago}
 \state{Illinois}
 \country{USA}}

\author{Richard Jean So}
\orcid{0009-0002-6442-055X}
\email{richard.so@duke.edu}
\affiliation{%
 \institution{Duke University}
 \city{Durham}
 \state{North Carolina}
 \country{USA}}
 
\author{Mina Lee}
\orcid{0000-0002-0428-4720}
\email{mnlee@uchicago.edu}
\affiliation{%
 \institution{University of Chicago}
 \city{Chicago}
 \state{Illinois}
 \country{USA}}

\renewcommand{\shortauthors}{Zhi et al.}

\begin{abstract}
\input{0.abstract}
\end{abstract}


\maketitle

\input{1.introduction}
\input{2.relatedwork}

\input{3.method}

\input{4.findings}

\input{5.behavioralengagement}
\input{6.discussion}
\input{7.conclusion}

\begin{acks}
We thank our participants for their time and the reviewers for their valuable feedback. This work was supported in part by the Social Science and Humanities Research Council (SSHRC) Insight Development Grant (IDG) for ``Literature in the Age of Artificial Intelligence'' (SSHRC 430-2024-00336). Any opinions, findings, conclusions, or recommendations expressed in this material are those of the authors and do not reflect the views of the funding agency.
\end{acks}

\bibliographystyle{ACM-Reference-Format}
\bibliography{main}

\appendix

\input{8.appendix}

\end{document}

%% file: 0.abstract.tex
AI demonstrates unprecedented reasoning capabilities, but its increasing integration into human reasoning via automated reading and summarization has provoked debate about its use for cultural interpretation. \emph{Close reading}---the practice of understanding, analyzing, and critiquing cultural texts for pleasure---is a skill at the core of such interpretation, traditionally being seen as exclusive to humans. To test AI's impact on close reading, both in terms of interpretative performance and pleasure, we conducted a preregistered randomized experiment ($n=400$) investigating the impact of AI assistance by presenting single or multiple AI interpretations, on close reading poems, compared to no AI assistance. We found that single AI interpretation boosted both performance and pleasure, while multiple AI interpretations only improved performance. Further exploration revealed a trade-off: participants who heavily relied on AI showed better performance on the task but lower pleasure. Our results contribute to discussion on whether and how to calibrate AI assistance for cultural interpretation: \textit{``less is more.''}

%% file: 1.introduction.tex


\section{Introduction}
The rise of Generative AI has provoked both optimism and fear around its impacts on human reasoning \cite{vaccaro2024combinations,singh2025protecting,tankelevitch2025understanding}. Some see opportunities for enhanced efficiency in consuming and creating information \cite{mollick2024co}, while others harbor concerns about cognitive outsourcing that bypasses human thought processes \cite{kosmyna2025brainchatgptaccumulationcognitive,gerlich2025ai,lee2025impact,oakley2025memory}. The debate revolves around the question of how much of human reasoning can or should be delegated to AI without making our own contributions irrelevant or unnecessary in the process.


While much of this debate has focused on the implications of AI for how we reason via writing \cite{lee2024empirical,lamiroy2022lamuse,mcguire2024establishing,rane2024role}, less attention has been given to the kinds of reasoning that happen via reading---especially when the materials are poems, songs, stories or movies, which demand focused and complex forms of interpretive attention. Recent work showed the benefits of large language models (LLMs) \emph{simplifying} large volumes of text to make it more accessible \cite{xiao2023inform,ash2024translating}, but do these same benefits accrue to our ability to \emph{interpret} culture and arts? And if the point of reading a poem or watching a movie is because you find pleasure in how it makes you think and feel, is there really much to be gained by having an LLM do it for you? 

This specific kind of interpretive skill, involving both attention and pleasure, is called \emph{close reading}: the ability to understand, explain, interpret, evaluate, and critique culture works \cite{abrams1999glossary,sui2025kristeva}---in textual form like poems and novels, or in other media like songs and films. Close reading is also foundational to the development of literacy and critical thinking, as it seeks to reveal non-obvious qualities about the cultural text through careful perception \cite{sinykin2025close,sui2025kristeva,hobbs2010digital,hayles2010we,krathwohl2002revision}. 

Though often associated with and widely taught within humanities education, people also practice close reading informally in everyday life, to varying degrees.
For instance, social media influencers rack up millions of followers by simply explaining popular TV shows and films \cite{screendaily2023influencers}. People seek to understand and interpret cultural texts on the Internet because doing so means in part understanding the social world itself better.
In this broader sense, close reading becomes not just an academic skill but also a marker of social engagement and cultural awareness \cite{geertz2017interpretation,carbaugh1991communication}.




The possibility that AI might automate close reading is controversial and has opened up a fierce debate among writers, creators, journalists, humanities professors, as well as common consumers of culture \cite{watkins2024ai,naquin2024close}. Few doubt that AI is a useful tool to automate instrumental tasks. But many question whether AI should be used to automate the task of interpretation---finding meaning in cultural texts---because it is typically imagined to be the thing that makes us human, one that cannot benefit from automation since its primary value lies in providing personal pleasure. Many fear that AI will somehow corrupt or diminish what is posited to be an exclusive human skill \cite{watkins2024ai}. As a matter of fact, there is already evidence of AI-generated interpretations being incorporated alongside cultural texts online, including poetry platforms \cite{pastan_love_poem}, making this investigation into AI's impact on close reading particularly timely and relevant to how cultural texts are actually being consumed in digital environments. 

To push the boundary of this discussion, this work seeks to better understand how close reading, and cultural interpretation in general, are affected by the integration of AI into our reading and reasoning. Given the importance of close reading as a social skill, our study focuses on lay readers,
and on poems as cultural texts. 
We investigate the impact of AI assistance powered by an LLM on close reading by examining two essential elements: \emph{interpretive performance} and the \emph{pleasure} derived from the process. We focus on identifying stylistic features and explaining their effects within the text, which is the first necessary step and foundation for effective close reading \cite{utaustin2024critical}, and operationalize interpretive performance through feature identification, interpretation quality, and writing quality. 
\revise{Building on close reading scholarship \cite{abrams1999glossary,guillory2025close,bialostosky2006should} and intrinsic rewards theory \cite{csikszentmihalyi2015intrinsic,csikszentmihalyi2014play}, we conceptualize the pleasure of close reading as arising from discovering personally resonant meanings, enjoying the interpretive puzzle-solving process, and feeling empowered to make sense of complex texts.} By this view, we operationalize these three sources of pleasure as three interrelated subjective experience constructs: appreciation, enjoyment, and self-efficacy \cite{abrams1999glossary, mcguire2024establishing}. 
With these two elements, we ask:


\smallskip
\hrule height 0.08em
\vspace{0.1em}
\begin{description}
    \item[RQ1.] How does exposure to AI assistance influence an individual’s interpretive performance at close reading \revise{poems}?
    \item[RQ2.] How does exposure to AI assistance influence the pleasure an individual derives from close reading \revise{poems}?
\end{description}
\vspace{0.1em}
\hrule height 0.08em
\smallskip


To answer these questions, we conducted a preregistered\footnote{https://aspredicted.org/555c-y7kz.pdf} randomized controlled experiment ($n=400$) with crowdworkers \revise{on Prolific}. 
We examined how different amounts of AI assistance influence individuals' interpretive performance and the pleasure they derive from close reading by comparing three conditions: (1) \singleai, which presented a single AI interpretation; (2) \multipleai, which offered multiple AI interpretations stacked on top of each other (i.e., shown one at a time, with the top interpretation fully visible); and (3) \noai, which did not provide any AI interpretation. \revise{The \singleai condition reflects the design used on poetry platforms, which display one AI interpretation beneath the poem \cite{pastan_love_poem}. The \multipleai condition tests the effect of providing more AI assistance by having additional interpretations accessible, aligning with the open-ended nature of close reading.}

In the study, each participant completed interpretation tasks for three poems in random order in their randomly-assigned condition. The interpretation tasks, adapted from the Critical Reader's Interpretive Toolkit (CRIT) \cite{utaustin2024critical}, focused on identifying stylistic features from the poems and analyzing their effects. 
After completing the task for each poem, they rated their appreciation and enjoyment of the poem, as well as their sense of self-efficacy in interpretation.


Our findings reveal that exposure to AI assistance in the form of a single AI interpretation boosted both performance and pleasure, while multiple AI interpretations improved performance but did not increase pleasure. 
\revise{Furthermore, when considering relative expertise (operationalized as experience of college-level humanities coursework), AI assistance enhanced pleasure only for inexperienced readers, while experienced readers showed no such benefits.} 
In further exploring how participants made use of their assigned AI assistance, based on behavioral engagement, we found that a considerable proportion of those exposed to multiple AI interpretations did not view all of them (42.1\% viewed one and 12.3\% viewed two out of three interpretations provided), and that a considerable proportion (54.8\% in \singleai, 47.0\% in \multipleai) of participants self-reported not having used the AI. 
These results and observations suggest, first and foremost, that modest exposure to AI assistance can provide benefits to the performance and pleasure of close reading, while presenting 
users with too much AI assistance can diminish these pleasurable benefits. 
Second, readers may naturally resist letting AI fully take over their interpretive work. 
Our data, analysis, and interface are available at the project website.\footnote{Our data, analysis, and interface code are available at \url{https://closereading-ai.app}, where readers can also explore the task interactively.}


Our work enriches the debates currently happening around AI and its consequences for human reasoning and cultural interpretation. Concretely, we make the following contributions: 1) We provide empirical evidence of how AI assistance can influence both interpretive performance and subjective experience of close reading, a typical human reasoning practice that many believe should not be assisted by AI; 
2) We demonstrate that modest exposure to AI assistance can be beneficial to both interpretive performance and pleasure, while extensive AI assistance provides some performance gains but may not help with pleasure; 3) We offer insights into how readers naturally regulate their use of AI assistance for close reading, 
showing resistance to having AI fully displace their interpretive work. From these findings, we highlight the need to carefully calibrate AI assistance in cultural interpretation to preserve the experiences we value as humans.


%% file: 2.relatedwork.tex
\section{Related Work}
\subsection{Close Reading and AI’s Capabilities of Interpretive Reasoning}

\looseness=-1 Scholarly traditions define close reading as the practice of analyzing ``the complex interrelations and ambiguities of verbal and figurative components'' within cultural works \cite{guillory2025close,abrams1999glossary,bialostosky2006should,coleridge1849notes}. It is a form of interpretive reasoning that centers on revealing how aesthetic choices in literary and cultural texts generate specific meanings through careful analysis of stylistic features \cite{abrams1999glossary,sui2025kristeva,bialostosky2006should}. What distinguishes this form of interpretive reasoning from problem-solving oriented reasoning tasks is its embrace of interpretive variety—rather than seeking singular correct solutions, close reading admits that cultural texts support numerous reasonable readings shaped by reader perspective and contextual factors \cite{sinykin2025close,bialostosky2006should,coleridge1849notes,farkas2015designing}. This quality fundamentally separates close reading from utility-driven reasoning tasks, positioning it instead as a form of cultural engagement where the process of interpretation itself generates value through enhanced understanding, and aesthetic appreciation and enjoyment \cite{guillory2025close,abrams1999glossary,hayles2010we}.

Close reading is taught in nearly every English high school class in North America and is central to college-level humanities course curricula \cite{carillo2018teaching,bialostosky2006should}. But people may develop this skill more or less given their education background and major. The significance of close reading extends beyond literary analysis, as educational research highlights that it is foundational to broader civic competencies of critical thinking for navigating information-rich digital environments today \cite{hobbs2010digital,hayles2010we,ebright2024we,krathwohl2002revision}. Research demonstrates that interpretive competencies developed through literary analysis transfer effectively to evaluating online sources, recognizing misinformation, and fostering informed civic participation in democracy \cite{mcgrew2020learning,ebright2024we,braun2020performance,shavelson2019assessment}.

The uniqueness of interpretive reasoning has attracted increasing attention among AI researchers \cite{crawford2021atlas,sui2025kristeva,stowe2022impli,tong2024metaphor}. Recent work has begun exploring LLM capabilities in domains requiring aesthetic judgment \cite{hullman2023artificial,yuan2024chatmusician}. Most notably, the KRISTEVA benchmark presents the first systematic evaluation of LLMs on close reading tasks, assessing models' abilities to identify stylistic features, and perform multi-hop reasoning between textual elements and external knowledge \cite{sui2025kristeva}. While these studies demonstrate the potential capabilities of state-of-the-art AI models for interpretive reasoning, little is known about how AI assistance leveraging such capabilities affects human close reading. To address this gap, our work examines how different forms of AI assistance affect both interpretative performance and subjective experiences of close reading. 


\subsection{Effects of AI on Reasoning via Reading}

With the increasing integration of AI into human reasoning, there is an ongoing debate over the benefits and harms of such integration. Much of this conversation has focused on the implications of AI for how we reason via \emph{writing}. For example, in creative writing, some studies demonstrated that AI assistance can enhance creative thinking by inspiring novel ideas \cite{lee2024empirical,lamiroy2022lamuse}, and other studies revealed that AI assistance may 
reduce content diversity \cite{padmakumar2023does,anderson2024homogenization}. 

However, we have less evidence about how the integration of AI impacts human reasoning that happens via \emph{reading}. Several recent works examine the impacts of LLMs on reading comprehension by assessing the correctness of human comprehension of the text. For example, \citet{kreijkes2025effects} demonstrated that the use of LLM combined with note-taking had significant positive effects on students’ comprehension of a text on history topics compared to the LLM alone. \citet{xiao2023inform} demonstrated that the integration of a purpose-built AI-powered chatbot can improve human comprehension of online consent forms. 
Yet we still know little about the impacts of LLMs when the material (e.g., poems, songs, stories, or movies) demands granular or complex forms of interpretive attention. Our work is different in that we focus on close reading tasks for cultural interpretation that require more interpretive attention than general reading comprehension, and admits no single correct answer as cultural texts support unlimited plausible interpretations contingent on subjective aesthetic judgment. 



The practical discourse around using AI for close reading reveals tensions between their transformative potential and possible risks. \citet{watkins2024ai} raised fundamental questions about whether AI tools, while enabling students to simplify complex academic texts and explore unfamiliar terms, encourage genuine engagement with texts or simply allow students to ``offload the burden of close reading.’’ \citet{naquin2024close} advocates for a ``three-part model’’ for close reading with LLMs that position AI as a cognitive augmentation tool rather than a replacement, emphasizing the symbiotic relationship between human critical thinking, the power of AI, and textual resources. These perspectives highlight the heated discussion and the necessity of empirical evidence about close reading with LLMs. Our study contributes empirical evidence to this debate by investigating the impact of AI, through expert-designed close reading tasks and rubrics adapted from humanities pedagogy given the challenge of measurement.

\subsection{Subjective Experience Measures in Human-AI Collaboration}

While experimental studies examining human-AI collaboration traditionally emphasizes performance metrics, recent studies in HCI have increasingly incorporated subjective experience measures that are meaningful for the context they focused on, revealing important tensions. \citet{wu2025human} measured intrinsic motivation, enjoyment, and perceived autonomy alongside task performance, finding that AI assistance decreased intrinsic motivation despite improving outcomes.  This ``double-edged sword'' effect suggests that AI assistance may diminish important experiential qualities despite performance gains. \citet{draxler2024ai} demonstrated that users rated their ownership of AI-assisted writing significantly lower than independent work. In creative domains, research has shown that AI tools can both inspire novel ideas and constrain authentic self-expression, with effects varying based on task characteristics and individual differences \cite{jia2024and,valenzuela2024artificial}. In educational contexts, studies have employed self-efficacy scales and learning motivation questionnaires \cite{yilmaz2023effect,bucinca2021proxy}, with mixed results. For example, \citet{yilmaz2023effect} found increased programming self-efficacy with ChatGPT use, while broader reviews suggest no overall effect on self-efficacy despite performance gains. 
Overall, these findings reveal that the directionality of effects on subjective experience measures can diverge from performance metrics, underscoring the need to examine not just what AI helps users accomplish, but how it affects their experience of the accomplishment itself.

\begin{figure*}[t]
\centering
\includegraphics[width=0.98\textwidth]{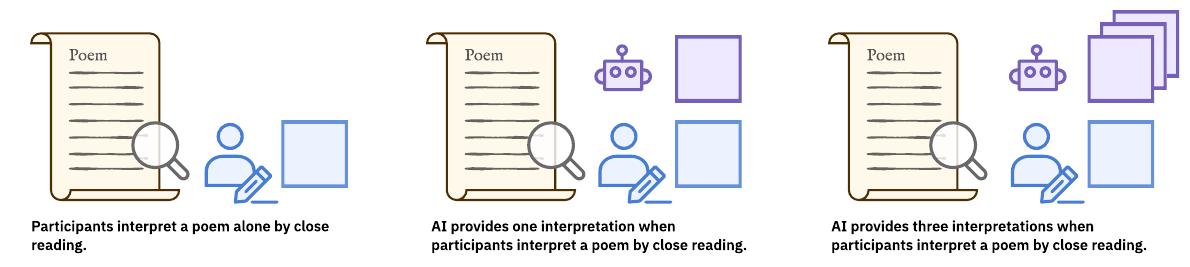}
\caption{Illustration of the three experimental conditions. Participants in all conditions interpret poems through close reading, with randomly assigned AI assistance: (left) \noai condition where participants interpret alone, (center) \singleai condition where one AI interpretation is provided, and (right) \multipleai condition where three AI interpretations are provided.}
\label{fig:conditions_diagram}
\end{figure*}

While performance metrics often take precedence in human-AI collaboration research, subjective experience deserves equal attention---particularly in domains like close reading where the experience is as valuable as the outcome. Unlike utility-driven tasks where efficiency gains represent clear benefits, close reading derives its value from the interpretive process itself. In our work, we measure three interrelated aspects of this subjective experience: Appreciation captures the personal resonance and emotional connection readers form with texts; Enjoyment reflects the intrinsic pleasure derived from the interpretive process; Self-Efficacy indicates readers' confidence in their interpretive abilities, which contributes to the subjective experience by fostering a sense of competence and mastery. Together, these aspects constitute what literary scholars recognize as the \emph{pleasure} of close reading \cite{sinykin2025close,hobbs2010digital,hayles2010we,krathwohl2002revision}---a multifaceted experience that encompasses both emotional satisfaction and intellectual confidence. We treat these subjective experience measures as equally important as performance metrics, recognizing that the ultimate goal of close reading is not merely to extract correct interpretations, but to foster meaningful encounters with literature that enrich readers' intellectual and emotional lives.

%% file: 3.method.tex
\section{Method}


\subsection{Study Design}

\subsubsection{Overview}

We designed a three-condition, between-subjects experiment ($n=400$) to investigate the effect of AI assistance on participants' interpretive performance and subjective experience of close reading. AI assistance was provided by presenting a single interpretation (\singleai) or multiple interpretations (\multipleai), compared to a control group with no AI assistance (\noai). Using a custom web interface (Figure \ref{fig:interface}), each participant performed close reading of three poems covering diverse topics. For each poem, participants performed three interpretation tasks identifying and explaining stylistic features, and answered questions about their subjective experience after interpretation. This process was repeated for all three poems in randomized order. Participants were randomly assigned to one of three conditions (\noai, \singleai and \multipleai; Figure \ref{fig:conditions_diagram}), applied consistently across all poems. 

\subsubsection{Conditions}\label{conditions}

\begin{figure*}[t]
\centering
\captionsetup[subfigure]{justification=raggedright, singlelinecheck=false}
\begin{subfigure}[t]{0.48\textwidth}
    \centering
    \includegraphics[width=\textwidth]{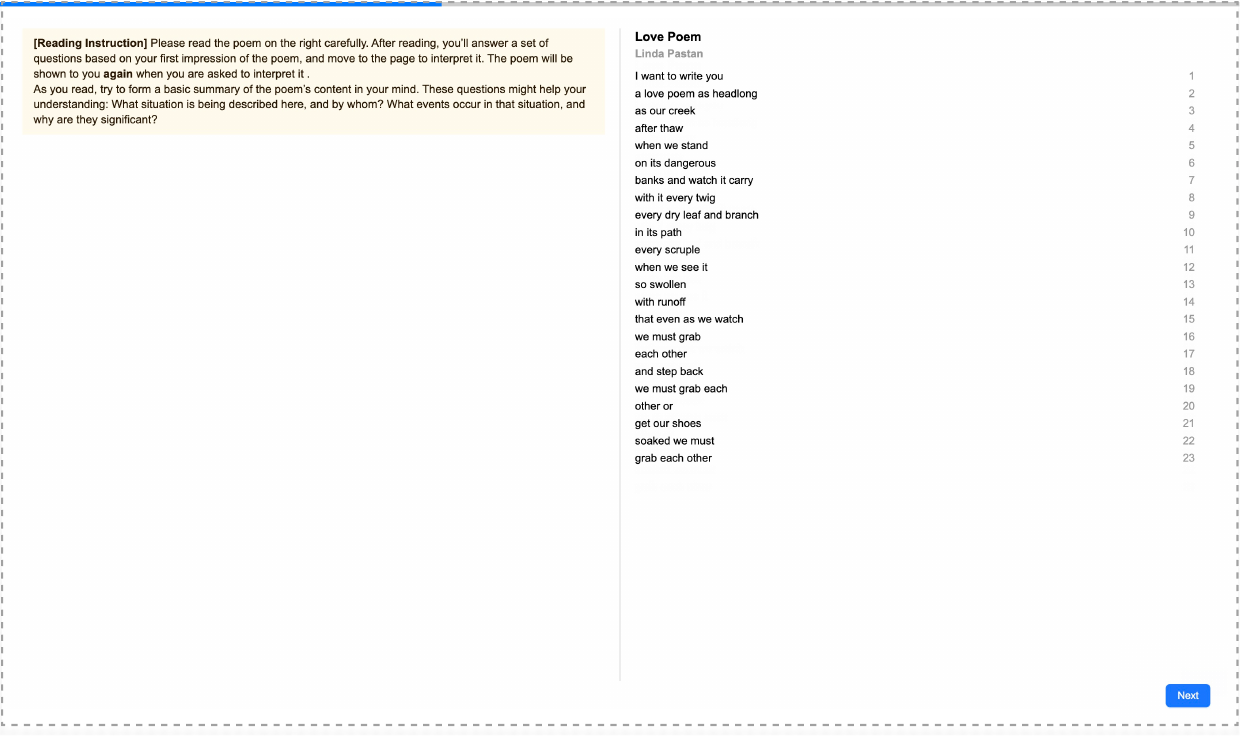}
    \caption{Initial reading of the poem before the interpretation task for all conditions.}
    \label{fig:interface_first_read}
\end{subfigure}
\hfill
\begin{subfigure}[t]{0.48\textwidth}
    \centering
    \includegraphics[width=\textwidth]{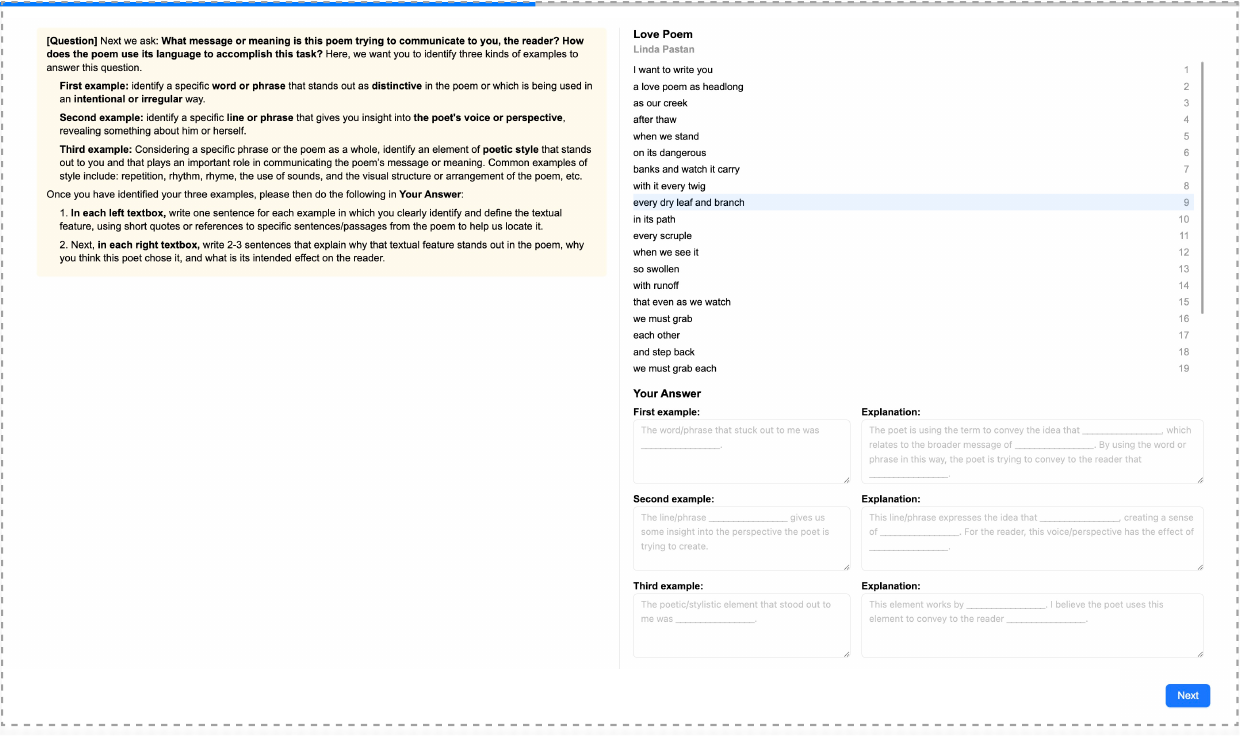}
    \caption{\noai: Close reading a poem without AI assistance.}
    \label{fig:interface_control}
\end{subfigure}
\vskip\baselineskip
\begin{subfigure}[t]{0.48\textwidth}
    \centering
    \includegraphics[width=\textwidth]{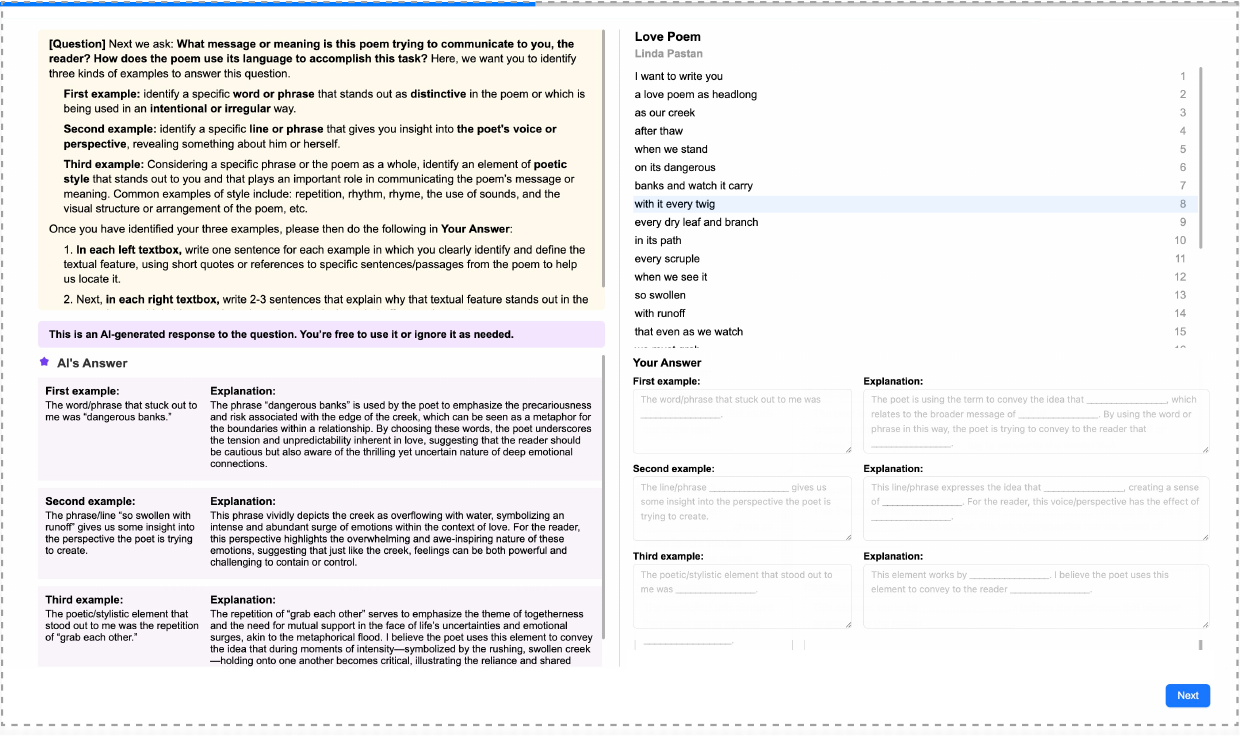}
    \caption{\singleai: Close reading a poem with one AI interpretation displayed.}
    \label{fig:interface_single}
\end{subfigure}
\hfill
\begin{subfigure}[t]{0.48\textwidth}
    \centering
    \includegraphics[width=\textwidth]{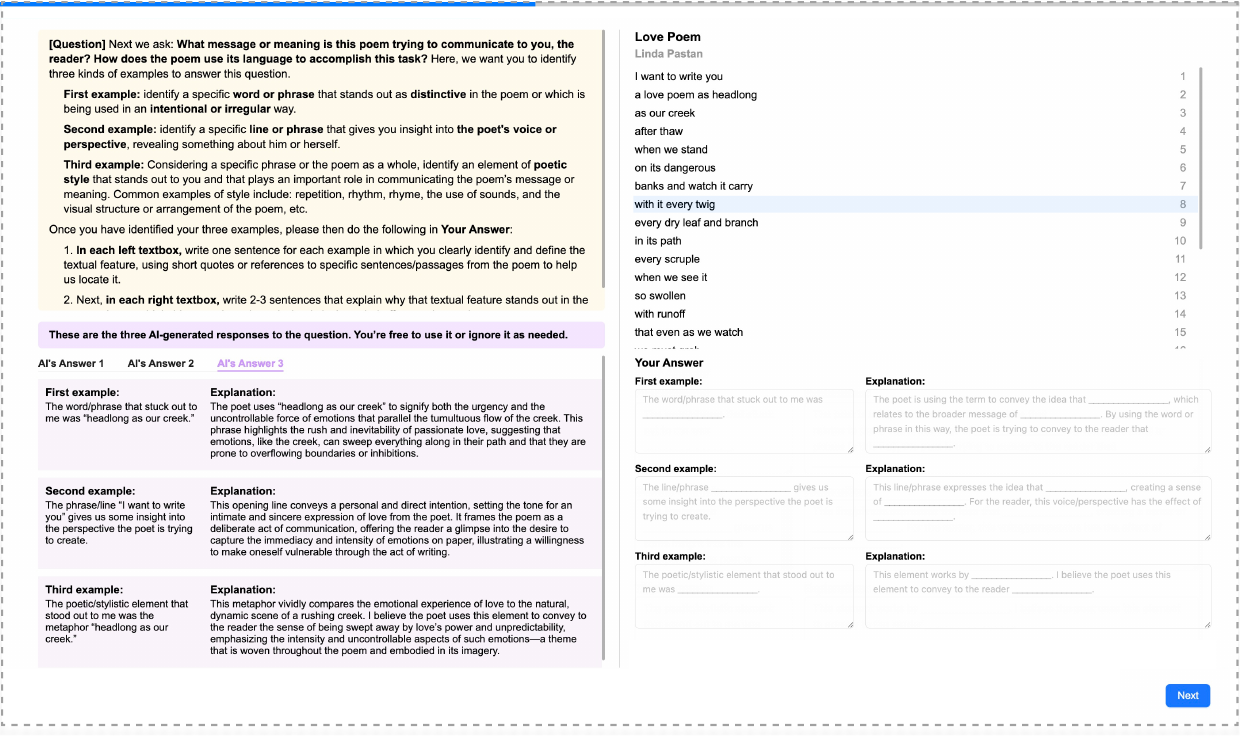}
    \caption{\multipleai: Close reading a poem with three AI interpretations stacked together, while one is displayed.}
    \label{fig:interface_multiple}
\end{subfigure}
\caption{Study interface showing the two-step process for each poem across experimental conditions. (a) All participants first read the poem briefly. (b-d) Participants then complete the close reading task, identifying and explaining three stylistic features with assigned AI assistance: no assistance in \noai, one AI interpretation provided in \singleai, or three AI interpretations provided in \multipleai. The two-step process follows the task structure in the CRIT \cite{utaustin2024critical}. The task structure remains identical across conditions; only the presence and amount of AI assistance varies.}
\label{fig:interface}
\end{figure*}

Participants were randomly assigned to one of three conditions:
\begin{itemize}
    \item \noai: Close reading a poem without assistance (Figure \ref{fig:interface_control}; Figure \ref{fig:conditions_diagram}, left).
    \item \singleai: Close reading a poem while being exposed to an AI interpretation (Figure \ref{fig:interface_single}; Figure \ref{fig:conditions_diagram}, center).
    \item \multipleai: Close reading a poem while being exposed to multiple AI interpretations (Figure \ref{fig:interface_multiple}; Figure \ref{fig:conditions_diagram}, right).
\end{itemize}
\revise{The \singleai condition reflects how poetry platforms may provide AI assistance alongside literary works \cite{pastan_love_poem}. The AI interpretation was displayed under the poem, visible to participants by default. The \multipleai condition tests a design choice of presenting more AI assistance. This responds to the open-ended nature of close reading, which allows for multiple valid interpretations. 
Three AI interpretations were available on a stacked interface. One was fully visible by default; two more were signaled by labeled buttons that participants could click to view.
We chose this stacked layout to reduce visual burden and give each AI interpretation equal visual space on the interface.}


\subsubsection{Interpretation Task and Poems}

For each poem, participants completed three interpretation tasks, each requiring them to identify a feature and explain its effect: (1) a distinctive word or phrase used intentionally or irregularly, (2) a line revealing the poet's voice or perspective, and (3) an element of poetic style (e.g., repetition, rhythm). For each feature, they explained in 2-3 sentences how it contributed to the poem's meaning or message.
These tasks and scoring rubrics were adapted from the Critical Reader's Interpretive Toolkit (CRIT) \cite{utaustin2024critical} and designed by domain experts in close reading education. The instruction of tasks was polished based on pilot studies (Pilot 1: $n=20$; Pilot 2: $n=60$; Pilot 3: $n=90$; Pilot 4: $n=100$) to ensure clarity and appropriate difficulty. The three interpretation tasks per poem provided a comprehensive assessment of participants' close reading performance (see B.2 in Supplementary Materials for detailed questions; B.4 for scoring rubrics).

We selected three poems through expert curation and iterative group discussion: \textit{``Love Poem''} by Linda Pastan \cite{pastan2005love}, \textit{``Dusting''} by Marilyn Nelson \cite{nelson1994dusting}, and \textit{``Theme for English B''} by Langston Hughes \cite{hughes2002theme}. Selection criteria prioritized readability for lay readers while ensuring diversity across multiple dimensions: themes and topics (love, nature, social identity, etc.), poetic styles, and author backgrounds. Our goal was to provide a representative selection of poems that lay readers as our participants could meaningfully engage, rather than to compare effects across different poems (see B.3 in Supplementary Materials for all the poem text).



\subsubsection{Study Procedure}
The study consisted of three phases: a pre-task survey, interpretation tasks for three poems, and a post-task survey (see B.1 in Supplementary Materials for detailed questions). The pre-task survey collected demographic information and participants' prior experience with college-level humanities coursework, which we used to classify their expertise in close reading. For each poem, participants followed a two-step process. First, they read the poem independently (Figure \ref{fig:interface_first_read}) and provided initial ratings of their first impression. Second, on the following page, they completed three interpretation tasks with or without AI assistance depending on their assigned condition (\noai: Figure \ref{fig:interface_control}; \singleai: Figure \ref{fig:interface_single}; \multipleai: Figure \ref{fig:interface_multiple}). After completing the interpretation tasks, participants rated their experience on three 7-point scales: Appreciation, ranging from ``I had a strongly negative reaction to it'' (1) to ``I had a strongly positive reaction to it'' (7); Enjoyment, ranging from ``Very unenjoyable'' (1) to ``Very enjoyable'' (7); and Self-efficacy, ranging from ``I feel strongly unable'' (1) to ``I feel strongly able'' (7). Participants also provided rationale for their ratings and reported whether they used the AI assistance (if provided). This process was repeated for all three poems, presented in randomized order. The post-task survey asked participants about their approach and rationale for using or not using the AI assistance provided, followed by overall study feedback.





\subsubsection{AI Interpretations Curated}

For the \singleai condition, we used a widely-used general purpose model at the time (GPT-4o) to generate one response for each poem using the exact instructions given to participants in the interpretation tasks. 
In other words, the LLM automated the same close reading task that participants were asked to complete. 
For the \multipleai condition, we prompted GPT-4o to generate three distinct interpretations to the same interpretation task instructions. \revise{Each AI interpretation identified different stylistic features and explained their effects accordingly.} 
To ensure quality, our graders evaluated all AI interpretations using the same rubrics applied to participants. 
The complete AI responses and prompts used to generate them are detailed in B.8 in Supplementary Materials.



\subsection{Dependent Variables} \label{DVs}
We measured two sets of dependent variables for each poem: Interpretative Performance and Subjective Experience measures.

\subsubsection{Interpretative Performance}
Participants completed three interpretation tasks per poem, designed by experts in close reading pedagogy. Two trained graders evaluated participants' response to each interpretation task based on the scoring rubric, assigning three types of scores for Interpretive Performance: 
\begin{itemize}
\item \textbf{Feature Identification}: Whether the identified feature was correct (0 = incorrect/missing, 1 = correct)
\item \textbf{Interpretation Quality}: Depth and insight of the explanation (1 = poor, 3 = average, 5 = excellent)
\item \textbf{Writing Quality}: Clarity and coherence of expression (1 = poor, 3 = average, 5 = excellent)
\end{itemize}
For analysis, we averaged each score type across the three tasks within each poem, yielding three Performance Measures (Feature Identification, Interpretation Quality, Writing Quality) per poem per participant.

\subsubsection{Subjective Experience}
After completing the interpretation tasks for each poem, participants rated their Subjective Experience on three 7-point Likert scales:
\begin{itemize}
\item \textbf{Appreciation}: Rating on the question ``To what degree did this poem resonate with you?'' (1 = strongly negative reaction, 7 = strongly positive reaction)     
\item \textbf{Enjoyment}: Rating on the question ``How much did you enjoy reading the poem?'' (1 = very unenjoyable, 7 = very enjoyable)
\item \textbf{Self-efficacy}: Rating on the question ``How confident are you in your ability to interpret the poem?'' (1 = strongly unable, 7 = strongly able)
\end{itemize}
Thus, we had three Subjective Experience Measures (Appreciation, Enjoyment, Self-efficacy) per poem per participant.

\subsection{Participants}
We recruited 405 participants from Prolific. We have 400 participants in all after excluding those who completed any of the poems in excessively short time (3 SDs below the mean). Based on a power analysis using pilot and simulated data, this sample size is needed for 80\% power, a medium effect size, using a significant level of 0.05. The overall study took around one hour to complete and each was paid £9.5. Participants had a 100\% approval rate on Prolific, based in the US, using English as their primary language and fluent in English. The study was approved by the Institutional Review Board (IRB) of our institution. 

Our randomized experiment yielded 400 participants in all: 141 in the \noai, 115 in the \singleai, and 14\revise{4} in the \multipleai. As each participant interpreted three poems, this means 423 instances in \noai, 345 instances in \singleai, and 43\revise{2} instances in \multipleai. \revise{A chi-squared test confirmed these differences were not statistically significant ($\chi^2(2) = 3.815$, $p = 0.149$).} 


\paragraph{\revise{Categorization of relatively experienced and inexperienced readers.}} We classified participants based on their relative expertise: 51.6\% had taken at least one college-level humanities course and were categorized as experienced readers, while the remaining 48.4\% were categorized as inexperienced.
\revise{This binary grouping, based solely on humanities coursework, provides only one indicator of close-reading expertise.}

\subsection{Analysis} \label{analysis}

\subsubsection{Grading Process}
To ensure reliable scoring of participants' interpretive responses, we employed two expert graders, both with undergraduate degrees in English and one a PhD student in English. The grading process began with extensive calibration, where graders completed 2-hour training sessions using pilot data to establish consensus on scoring criteria and interpretation of the rubrics. To assess inter-rater reliability, we stratified a validation set of 70 participants (17.5\% of the total sample) drawn equally from all three experimental conditions. Both graders independently scored all responses in this validation set. For Feature Identification accuracy, the agreement between graders was 85.4\%. For Interpretation Quality and Writing Quality, the correlation coefficient ICC(2,1) is 0.76, exceeding the recommended threshold of 0.70 \cite{gisev2013interrater,hallgren2012computing}.

\subsubsection{Statistical Analysis}
Following our pre-registration\footnote{https://aspredicted.org/555c-y7kz.pdf}, for our main analyses, we employed mixed-effects models to analyze our data. For Performance Measures (Feature Identification Accuracy, Interpretation Quality, and Writing Quality), we fitted mixed-effects linear regression models with conditions (\noai, \singleai, \multipleai), expertise (inexperienced versus experienced readers), poem types, poem positions as fixed effects, and participant as a random effect. For Subjective Experience Measures (Appreciation, Enjoyment, and Self-Efficacy), we used mixed-effects ordinal regression models with the same fixed and random effects structure. 




%% file: 4.findings.tex
\input{Tables/linear_regression_table}

\begin{figure*}[!htbp]
\centering
\begin{subfigure}{\textwidth}
    \centering
    \includegraphics[width=0.95\textwidth]{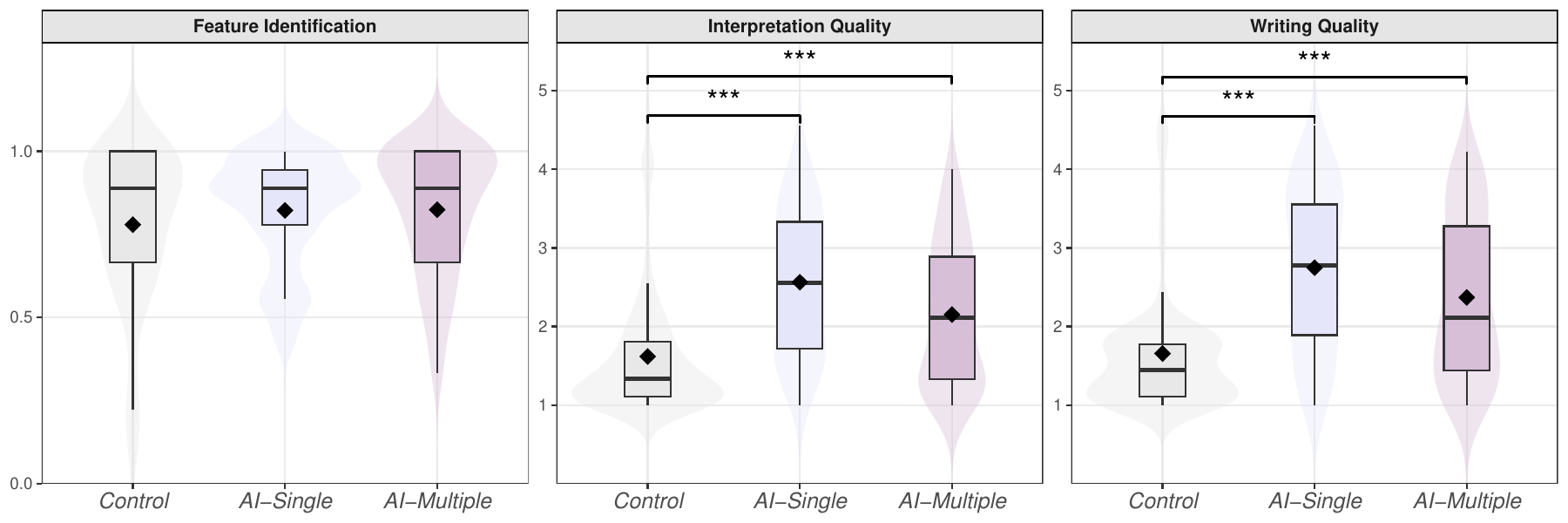}
    \caption{Inexperienced readers: The \singleai and \multipleai conditions demonstrated significant improvement of Interpretation Quality and Writing Quality compared to the \noai.}
    \label{fig:novice_performance_sub}
\end{subfigure}
\\[1em] 
\begin{subfigure}{\textwidth}
    \centering
    \includegraphics[width=0.95\textwidth]{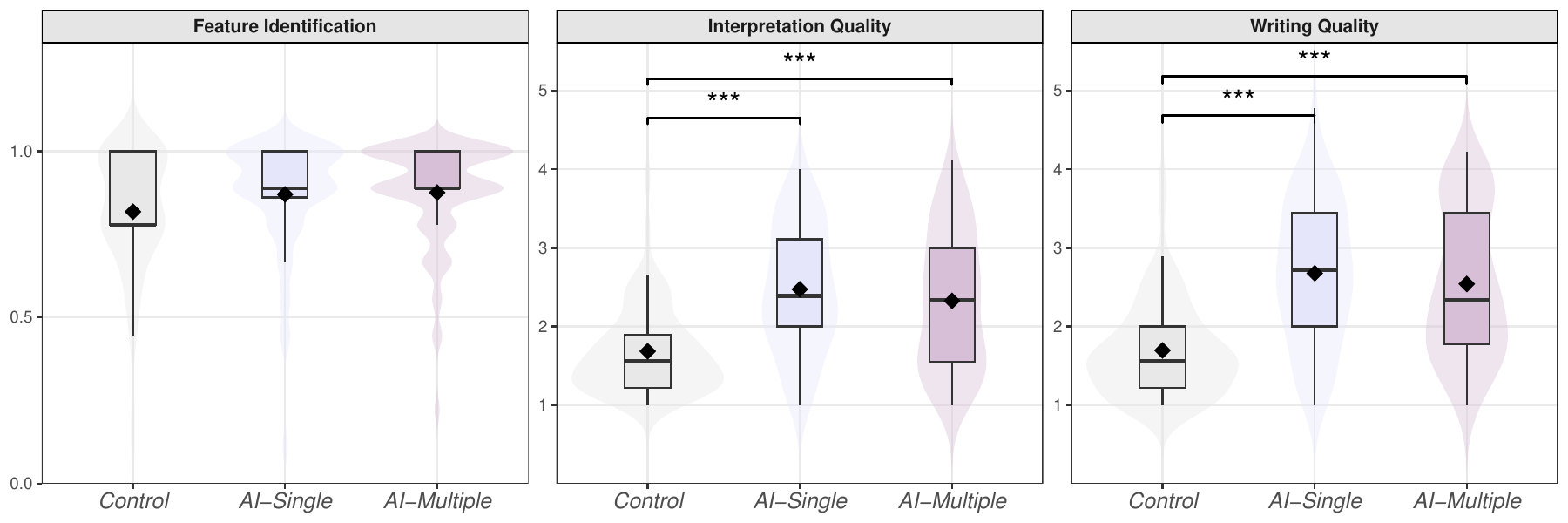}
    \caption{Experienced readers: The \singleai and \multipleai conditions demonstrated significant improvement of Interpretation Quality and Writing Quality compared to the \noai.}
    \label{fig:expert_performance_sub}
\end{subfigure}
\caption{Comparison of \textbf{Interpretive Performance} measures (Feature Identification, Interpretation Quality, and Writing Quality) between \noai and \singleai, and between \noai and \multipleai within (a) inexperienced and (b) experienced readers. In each plot, the diamonds represent means, the box plots represent medians and interquartile ranges, and the violin plots represent distributions. Significance differences are annotated based on adjusted p-values after correction ( * $<0.05$, ** $<0.01$, *** $<0.001$).}
\label{fig:combined_expertise_performance}
\end{figure*}

\section{Results}

In this section, we present the results of our pre-registered main analyses examining the impact of AI assistance on participants’ close reading performance and the pleasure derived from the process. We begin with how AI assistance affected participants' Interpretive Performance (Section~\ref{RQ1_results}). Then, we present the results of how AI assistance influenced their Subjective Experience (Section~\ref{RQ2_results}).

\subsection{Interpretive Performance (RQ1)}\label{RQ1_results}

As introduced in Section~\ref{DVs}, participants' interpretations of each poem were evaluated across three measures capturing different aspects of close reading proficiency: Feature Identification (scored 0-1), Interpretation Quality (scored 1-5), and Writing Quality (scored 1-5). Overall, we found that:

\begin{itemize}
\mainfinding{1}{AI assistance consistently improved participants' close reading performance across all measures.}
\mainfinding{2}{AI assistance by presenting a single interpretation showed larger effect sizes than multiple interpretations.}
\mainfinding{3}{Both inexperienced and experienced readers benefited from AI assistance in their close reading performance.}
\end{itemize}

\subsubsection{The Effect of AI Assistance}\label{effect_AI_performance}

Table \ref{tab:linear_regression_results} shows a consistent pattern that both \singleai and \multipleai significantly enhanced all performance measures (Feature Identification, Interpretation Quality, and Writing Quality). Compared to the \noai condition, the \singleai condition showed a significant positive effect on Feature Identification (0.048, 95\% CI: [0.002, 0.094], $p = 0.042$), Interpretation Quality (0.865, 95\% CI: [0.664, 1.065], $p < 0.001$), and Writing Quality (1.035, 95\% CI: [0.817, 1.253], $p < 0.001$). \revise{Note that these coefficients represent the estimated score difference compared to the \noai: for example, a coefficient of 0.865 for \singleai on Interpretation Quality means participants in this condition scored 0.865 points higher on average than those in the \noai.}
The \multipleai condition also showed significant positive effects compared to \noai across all performance measures: Feature Identification (0.051, 95\% CI: [0.008, 0.095], $p = 0.020$), Interpretation Quality (0.593, 95\% CI: [0.404, 0.782], $p < 0.001$), and Writing Quality (0.784, 95\% CI: [0.578, 0.989], $p < 0.001$). It is worth noting that the \singleai condition showed larger effect sizes than \multipleai across all measures.


Experienced readers showed a small but significant advantage in Feature Identification (0.046, 95\% CI: [0.010, 0.083], $p = 0.014$), but no significant differences in Interpretation Quality ($p = 0.454$) or Writing Quality ($p = 0.534$). While experienced readers were better at concisely identifying stylistic features in poems, this advantage did not extend to explaining the effects of these features or to the writing quality of their explanations. This suggests that when inexperienced readers are supported by AI, they may achieve comparable performance to experienced readers in these complex analytical and communicative tasks.


\subsubsection{Did the Effects of AI Assistance on Interpretation Performance Vary Among Inexperienced and Experienced Readers?}\label{interpretation_performance_by_expertise}


Since all participants completed the same interpretation tasks for the same three poems, we compare the average performance across all poems between \noai vs. \singleai, and \noai vs. \multipleai within each expertise group using Mann-Whitney U tests, \revise{with p-values Holm-Bonferroni adjusted for multiple comparisons.}

As shown in Figure \ref{fig:combined_expertise_performance}, both inexperienced and experienced readers benefited from both forms of AI assistance in their Interpretive Performance, with significant improvements in Interpretation and Writing Quality, regardless of prior development of close reading skills. Table \ref{tab:expertise_group_performance} shows that both forms of AI assistance significantly improved Interpretation and Writing Quality for both inexperienced and experienced readers (all $p.adj < 0.001$). Feature Identification exhibited minimal improvement, with only experienced readers in the \multipleai condition showing a marginally significant ($p.adj = 0.059$) but small effect, echoing the small effect sizes for Feature Identification observed in Table \ref{tab:linear_regression_results}.

\input{Tables/ordinal_regression_table}

\begin{figure*}[!htbp]
\centering
\begin{subfigure}{\textwidth}
    \centering
    \includegraphics[width=0.95\textwidth]{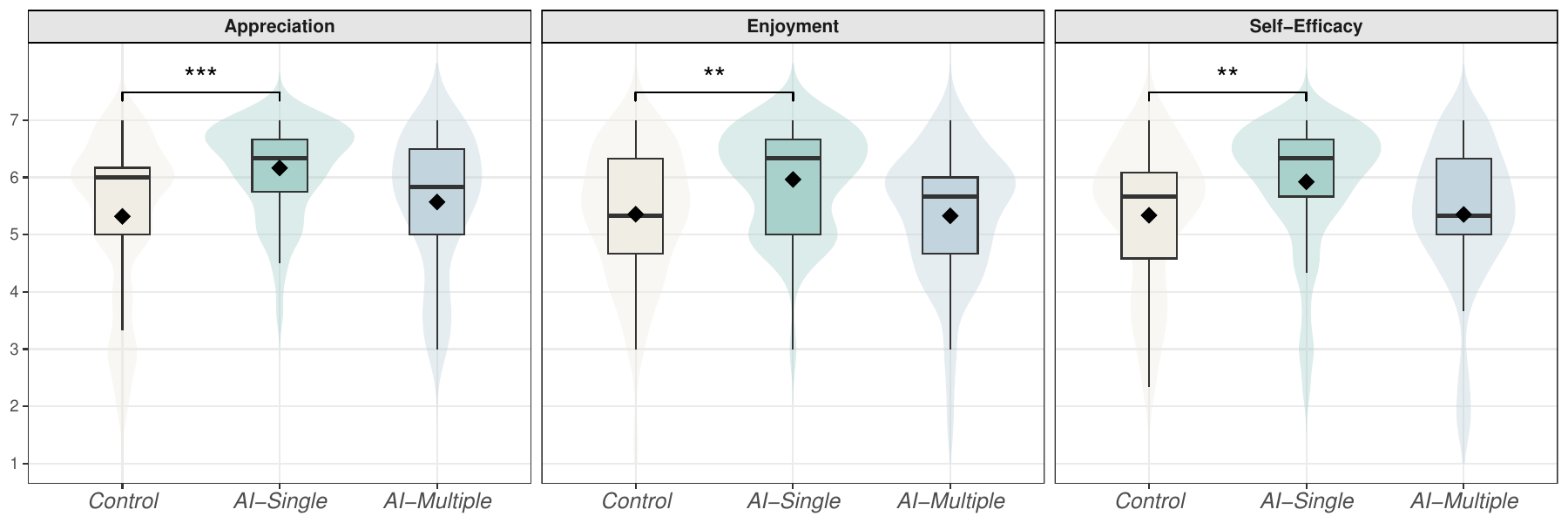}
    \caption{Inexperienced Readers: Those in the \singleai condition rated all three measures significantly higher than those in the \noai condition. No significant difference found between the \noai condition and the \multipleai condition.}
    \label{fig:novice_sub}
\end{subfigure}
\\[0.5em] 
\begin{subfigure}{\textwidth}
    \centering
    \includegraphics[width=0.95\textwidth]{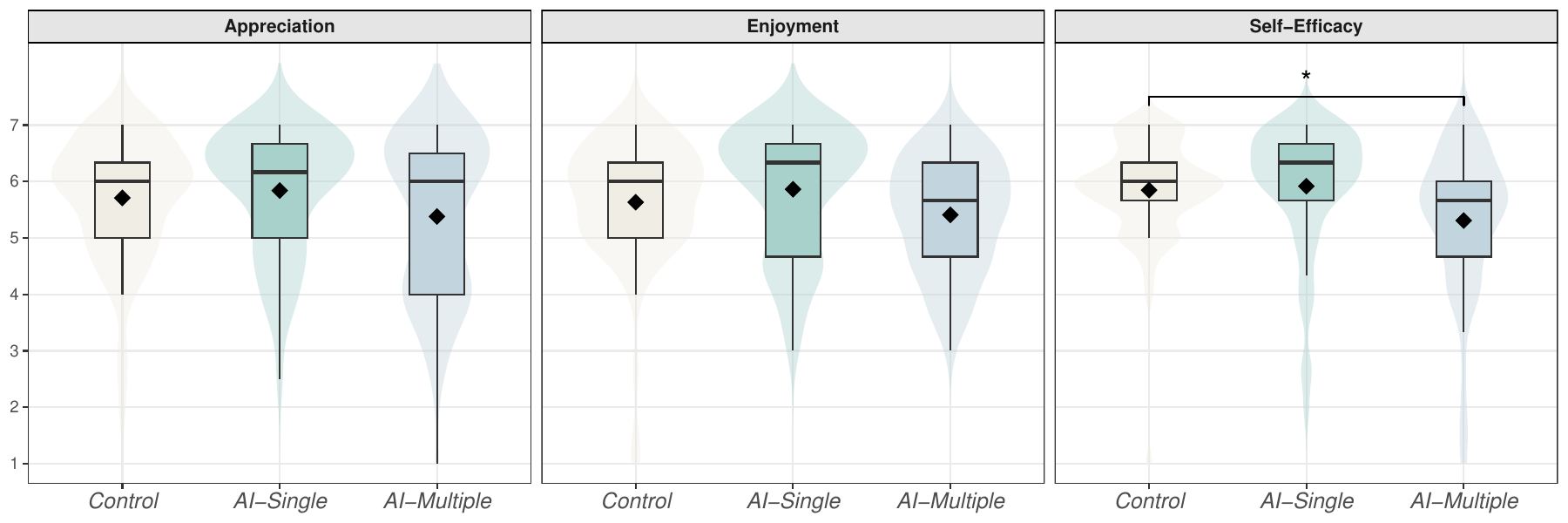}
    \caption{Experienced Readers: No significant difference found between the \noai condition and the \singleai condition. Those in the \multipleai condition rated Self-Efficacy significantly lower than those in the \noai condition.}
    \label{fig:expert_sub}
\end{subfigure}
\caption{Comparison of \textbf{Subjective Experience} measures (Appreciation, Enjoyment, and Self-efficacy) between \noai and \singleai, and between \noai and \multipleai within (a) inexperienced and (b) experienced readers. In each plot, the diamonds show means, the box plots show medians and interquartile ranges, and the violin plots show distributions. Significance differences are annotated based on adjusted p-values after correction ( * $<0.05$, ** $<0.01$, *** $<0.001$).}
\label{fig:combined_expertise}
\end{figure*}

\subsection{Pleasure Derived From Interpretation (RQ2)}\label{RQ2_results}

After close reading each poem, participants rated their Appreciation, Enjoyment, and Self-Efficacy of interpreting the poem on a 7-point likert scale. Overall, we found that:

\begin{itemize}
\mainfinding{4}{AI assistance by presenting a single interpretation improved participants' Enjoyment, Appreciation, and Self-Efficacy, while AI assistance by presenting multiple interpretations showed no benefits.}
\mainfinding{5}{Breaking this down further, only inexperienced readers benefited from AI assistance in the form of a single interpretation for their Enjoyment, Appreciation, and Self-Efficacy.}
\mainfinding{6}{Experienced readers showed no benefits from AI assistance for Subjective Experience. In fact, AI assistance in the form of multiple interpretations reduced their Self-Efficacy.}
\end{itemize}

\subsubsection{The Effect of AI Assistance}

Table \ref{tab:ordinal_regression_results} presents a consistent pattern across Enjoyment, Appreciation, and Self-Efficacy: AI assistance by presenting a single interpretation consistently enhanced all three aspects related to pleasure, while AI assistance by presenting multiple interpretations had no statistical significance. 
Compared to the \noai condition, the \singleai condition showed a significant positive effect on Appreciation ratings (0.899, 95\% CI: [0.522, 1.277], $p < 0.001$), Enjoyment ratings (0.969, 95\% CI: [0.531, 1.407], $p < 0.001$), and Self-Efficacy ratings (0.90, 95\% CI: [0.356, 1.443], $p = 0.001$). 
\revise{Note that these coefficients are log-odds from ordinal regression, where a positive value indicates higher odds of reporting higher ratings: for example, a coefficient of 0.969 for \singleai on Enjoyment means that participants in this condition were 2.64 times more likely to report higher Enjoyment ratings than those in the \noai.}
There was no significant difference between the \multipleai compared to the \noai condition.

\subsubsection{Did AI Assistance Influence the Pleasure of Close Reading Among Inexperienced and Experienced Readers Differently?}

As all participants rated their pleasure after the same interpretation tasks for the same three poems, we also compare the average Appreciation, Enjoyment, and Self-efficacy ratings across all poems between \noai vs. \singleai and \noai vs. \multipleai within expertise groups. 



As shown in Figure \ref{fig:novice_sub}, inexperienced readers in the \singleai condition showed significant improvements across all three subjective experience measures compared to \noai (all $p.adj < 0.01$), with moderate effect sizes. In contrast, the \multipleai condition showed no significant differences from \noai for any Subjective Experience measure. See Table \ref{tab:subjective_measures} for the detailed effect sizes and p-values. 
For experienced readers (Figure \ref{fig:expert_sub}), neither form of AI assistance improved these Subjective Experience measures. Notably, the \multipleai condition significantly reduced Self-Efficacy compared to \noai ($p.adj = 0.021$, $r = -0.257$), suggesting that multiple AI interpretations may undermine experienced readers' confidence in their interpretive abilities.

\vspace{2em}

Overall, results of our pre-registered main analyses revealed that while AI assistance by presenting a single interpretation and AI assistance by presenting multiple interpretations both improved participants' interpretive performance. Only AI assistance by presenting a single interpretation increased participants' pleasure. To better understand these results, in the next section, we further investigate how participants engaged with AI assistance using descriptive analyses of both behavioral logs and self-reported data.


%% file: Tables/linear_regression_table.tex
\begin{table*}[!htbp]
\centering
\small
\begin{tabular}{@{}lp{0.21\textwidth}p{0.22\textwidth}p{0.21\textwidth}@{}}
\toprule
& \textbf{Feature Identification} & \textbf{Interpretation Quality} & \textbf{Writing Quality} \\
\midrule
\textbf{Condition (ref: \noai):} & & & \\
\rowcolor{blue!8}\quad \singleai & 0.048* [0.002, 0.094] & 0.865*** [0.664, 1.065] & 1.035*** [0.817, 1.253] \\
\rowcolor{blue!8}\quad \multipleai & 0.051* [0.008, 0.095] & 0.593*** [0.404, 0.782] & 0.784*** [0.578, 0.989] \\
\textbf{Poem Position (ref: Position 1):} & & & \\
\quad Position 2 & 0.017 [-0.009, 0.043] & 0.006 [-0.059, 0.071] & 0.040 [-0.022, 0.102] \\
\quad Position 3 & 0.010 [-0.015, 0.036] & 0.084* [0.019, 0.149] & 0.111*** [0.049, 0.173] \\
\textbf{Poem Type (ref: Love Poem):} & & & \\
\quad Dusting & -0.003 [-0.029, 0.023] & -0.017 [-0.082, 0.048] & -0.024 [-0.086, 0.038] \\
\quad Theme for English B & 0.074*** [0.048, 0.100] & -0.110*** [-0.175, -0.045] & -0.089** [-0.151, -0.027] \\
\textbf{Participant Expertise (ref: Inexperienced):} & & & \\
\quad Experienced & 0.046* [0.010, 0.083] & 0.061 [-0.099, 0.220] & 0.055 [-0.118, 0.228] \\
\midrule
\end{tabular}
\begin{minipage}{\textwidth}
\small
\textit{Note:} + $p < 0.1$, * $p < 0.05$, ** $p < 0.01$, *** $p < 0.001$.
\end{minipage}\\[-0.6em]
\noindent\rule{\textwidth}{0.6pt}
\caption{Mixed-effects linear regression models predicting \textbf{Interpretive Performance} measures (Feature Identification, Interpretation Quality, and Writing Quality) from conditions, poems and participant expertise. We report coefficients with 95\% confidence intervals in brackets and highlight significant effects that are consistent on all the three performance measures.}
\label{tab:linear_regression_results}
\end{table*}

%% file: Tables/ordinal_regression_table.tex
\begin{table*}[!htbp]
\centering
\small
\begin{tabular}{@{}lp{0.21\textwidth}p{0.22\textwidth}p{0.21\textwidth}@{}}
\toprule
& \textbf{Appreciation} & \textbf{Enjoyment} & \textbf{Self-Efficacy} \\
\midrule
\textbf{Condition (ref: \noai):} & & & \\
\rowcolor{blue!8}\quad \singleai & 0.899*** [0.522, 1.277] & 0.969*** [0.531, 1.407] & 0.900** [0.356, 1.443] \\
\quad \multipleai & -0.070 [-0.416, 0.276] & -0.248 [-0.645, 0.149] & -0.352 [-0.856, 0.152] \\
\textbf{Poem Position (ref: Position 1):} & & & \\
\quad Position 2 & -0.127 [-0.384, 0.130] & -0.042 [-0.303, 0.219] & -0.163 [-0.436, 0.110] \\
\quad Position 3 & -0.096 [-0.356, 0.164] & 0.010 [-0.256, 0.276] & -0.118 [-0.393, 0.157] \\
\textbf{Poem Type (ref: Love Poem):} & & & \\
\quad Dusting & -0.125 [-0.379, 0.129] & -0.360** [-0.623, -0.097] & -0.196 [-0.471, 0.079] \\
\quad Theme for English B & 0.802*** [0.537, 1.067] & 0.431** [0.164, 0.698] & 0.199 [-0.073, 0.472] \\
\textbf{Participant Expertise (ref: Inexperienced):} & & & \\
\quad Experienced & -0.114 [-0.409, 0.181] & 0.148 [-0.191, 0.487] & 0.191 [-0.237, 0.619] \\
\midrule
\end{tabular}
\begin{minipage}{\textwidth}
\small
\textit{Note:} + $p < 0.1$, * $p < 0.05$, ** $p < 0.01$, *** $p < 0.001$.
\end{minipage}\\[-0.6em]
\noindent\rule{\textwidth}{0.6pt}
\caption{Mixed-effects ordinal regression models predicting \textbf{Subjective Experience} measures (Appreciation, Enjoyment, and Self-Efficacy) from conditions, poems and participant expertise. We report coefficients as log-odds with 95\% confidence intervals in brackets and highlight significant effects that are consistent on all the three subjective experience measures.}
\label{tab:ordinal_regression_results}
\end{table*}

%% file: 5.behavioralengagement.tex
\section{Behavioral Engagement}\label{behavioral_engagement}



In this section, we report observations on the following: participants’ viewing behavior of AI interpretations in the \multipleai condition, as limited engagement might explain the lack of pleasurable benefits (Section \ref{viewing_AI_multiple}); time allocation of activities during the task across conditions (Section \ref{time_allocation}); self-reported AI use (Section \ref{self_reported_AI_use}) and textual overlap between participant and AI interpretations (Section \ref{copying_AI}) in the \singleai and \multipleai conditions. Together, these observations provide a comprehensive picture of how participants used, and perceived their use of, the AI assistance to which they were exposed. We summarize our observations as follows:

\begin{itemize}
    \item \textbf{Limited Engagement with Multiple AI Interpretations:} Despite having access to three AI interpretations, 54.4\% of participants in \multipleai chose not to view all available options (42.1\% viewed only one; 12.3\% viewed two). Participants who viewed multiple interpretations frequently reported feelings of inadequacy and competitive pressure.
    \item \textbf{Self-Reported AI Use Patterns:} A proportion of participants reported not using AI assistance despite its availability (\singleai: 45.2\%; \multipleai: 53.0\%). 
    \item \textbf{Performance-Pleasure Trade-off:} Participants who reported using AI or whose responses showed high textual overlap with AI interpretations achieved performance scores closer to the AI benchmark but consistently reported lower subjective experience ratings across enjoyment, appreciation, and self-efficacy measures.
\end{itemize}

\subsection{Did Participants View the Multiple AI Interpretations?}\label{viewing_AI_multiple}

In the \multipleai condition, participants were shown three AI interpretations stacked together with one fully visible by default and two more signaled by labeled buttons that participants could click to view. Table \ref{tab:viewing_patterns} shows that participants viewed only the default AI interpretation in 42.1\% of instances, clicked to view a second AI interpretation in 12.3\% of instances, and clicked through all three in 45.6\% of instances, suggesting limited engagement with the multiple AI interpretations. While participants who viewed multiple AI interpretations showed better Interpretive Performance in Interpretation Quality and Writing Quality, they had lower Subjective Experience in Enjoyment and Self-Efficacy compared to those who viewed only one AI interpretation in the \multipleai or those in the \singleai and \noai. 
Notably, those who viewed only one AI interpretation in the \multipleai showed lower Interpretive Performance and Subjective Experience, compared to those in the \singleai. 
\revise{Results of exploratory analyses aligned with this: after adjusting for the number of AI interpretations viewed, participants in the \multipleai showed lower Interpretive Performance and Subjective Experience than \singleai (see B.1 in Supplementary Materials). This pattern suggests that the mere visible presence of multiple AI interpretations may have negative effects.}



The availability of multiple AI interpretations did not translate into a better subjective experience for participants, as one might expect given the wider selection of possible interpretations. Instead, multiple AI interpretations might trigger feelings of inadequacy and competition, harming the subjective experience. Those who viewed multiple interpretations might feel discouraged by the AI's capabilities, as P27 in the \multipleai noted \revise{in their open-ended responses reflecting on their use of the AI assistance}: \textit{``All the answers by the AI again interpreted the poem much better than I could've. I only had a surface level interpretation. [...]''} Multiple AI interpretations could create pressure to compete, and many chose not to view additional interpretations to preserve their self-confidence. P36 in the \multipleai explained this pressure and preservation strategy: \textit {``I was tempted to read the other ones to make sure we hadn't come up with the same examples. I felt like the AI was a threat that I had to do better than. However, I rationalized that if I didn't read what it wrote at all, then I would know for certain that it had not influenced my interpretation whatsoever.''}

The limited engagement with and reluctance to explore multiple AI interpretations may explain the absence of pleasurable benefits for \multipleai. Moreover, since each participant interpreted three poems, feelings from viewing multiple interpretations on one poem potentially carried over to subsequent poems, affecting the overall experience even when participants chose not to view multiple interpretations later.

\subsection{How Did Participants Allocate Their Time During the Task?}\label{time_allocation}

We checked the time allocation of participants' cursor movements across conditions to understand their attention levels during the task. These observations should be interpreted as suggestive patterns of how participants distributed their cognitive resources during the task. Cursor events can be noisy and provide only a heuristic measure of attention allocation, as participants may read or think about content without their cursor over it.

\begin{figure*}[!htbp]
\centering
\includegraphics[width=0.95\textwidth]{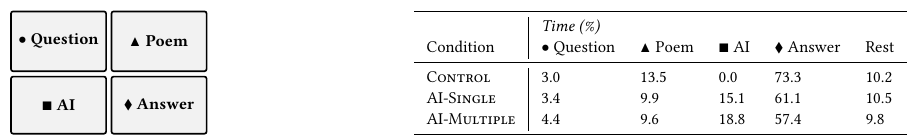}
\caption{Left: Layout of the four areas during the poetry interpretation task in the study interface. Right: Averaged percentage of time participants spent with their cursor in each area. The data shows how participants potentially allocated their attention across the question area (interpretation instructions), poem area (poem text), AI area (not available in Control), and answer area (participant answer fields) during the task.}
\label{fig:cursor_activity}
\end{figure*}

Figure \ref{fig:cursor_activity} shows that participants in \multipleai spent the highest percentage of time in the AI area (18.8\%) but the lowest percentage in the Poem area (9.6\%) among all conditions. Combined with our earlier findings that multiple AI interpretations triggered feelings of inadequacy and competition, this time allocation potentially suggests cognitive overload in that participants spent more time processing AI interpretations that did not help their subjective experience. While participants in the \multipleai spent the highest percentage of time in the AI area, the difference in the percentage of time spent in the AI area between the \singleai condition and \multipleai condition was moderate, echoing our earlier finding of limited participation with multiple interpretations.

\begin{table*}[!htbp]
\centering
\includegraphics[width=0.9\textwidth]{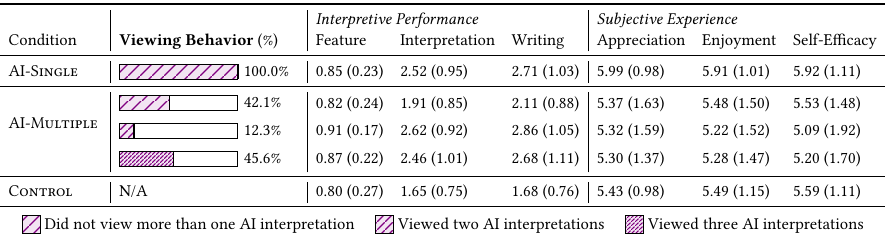}
\caption{Percentage of instances viewing different numbers of AI interpretations in the \multipleai condition, and summaries of Interpretive Performance and Subjective Experience measures (means with standard deviations in brackets). The \singleai and \noai are included for reference.}
\label{tab:viewing_patterns}
\end{table*}

\subsection{Did Participants Report Using AI?}\label{self_reported_AI_use}

After interpreting each poem, participants responded to a question about whether they used the AI for interpretation. Table \ref{tab:self_reported_ai_use} shows that a proportion of participants reported not using AI assistance despite its availability (\singleai: 45.2\%; \multipleai: 53.0\%). 

Within the \singleai or \multipleai condition, participants who reported using AI demonstrated better Interpretive Performance but consistently had lower Subjective Experience compared to those who did not report AI use. This pattern reflects a performance-pleasure tradeoff: those who acknowledged using AI showed better analytical output but diminished subjective rewards. Notably, in \singleai, participants who reported not using AI had the highest Subjective Experience of any group.

Comparing across conditions, among those who reported using AI, participants in \singleai had higher Subjective Experience and larger Interpretive Performance gains than their \multipleai counterparts, who showed the lowest Subjective Experience across all groups. Among those who did not report using AI, participants in \singleai similarly outperformed those in \multipleai on both Subjective Experience and Interpretive Performance. \revise{This suggests that, regardless of how participants characterized their AI use, offering multiple AI interpretations might not provide additional benefits over a single interpretation.}
Notably, participants in both \singleai and \multipleai who did not report using AI showed better Interpretive Performance than those in \noai, yet only those in \singleai showed better Subjective Experience than \noai. \revise{This pattern aligns with our main analyses and suggests that participants may have learned from the AI implicitly, even when they did not explicitly acknowledge ``using'' it.}


Furthermore, Table \ref{tab:ai_use_expertise_2} shows that, in \singleai, a lower percentage of experienced readers reported using AI, in contrast to inexperienced readers. However, in the \multipleai, the percentage of self-reported AI use among experienced readers increased. This suggests that while experienced readers may have strong resistance to using AI for interpreting poems when exposed to a single interpretation, they were more open to using it when given multiple AI interpretations.

\begin{table*}[!htbp]
\centering
\includegraphics[width=0.9\textwidth]{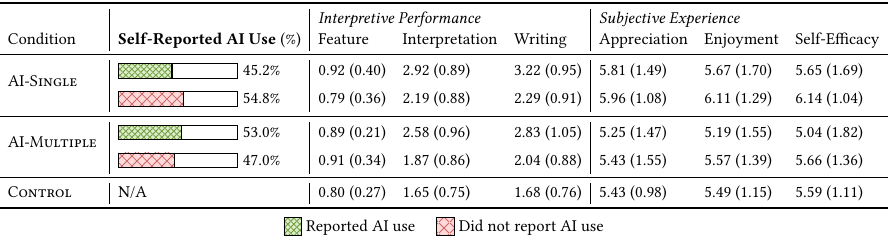}
\caption{Percentage of instances of self-reported AI use in each condition, and summaries of Interpretive Performance and Subjective Experience measures (means with standard deviations in brackets). The \noai is included for reference.}
\label{tab:self_reported_ai_use}
\end{table*}

\begin{table*}[!htbp]
\centering
\includegraphics[width=0.6\textwidth]{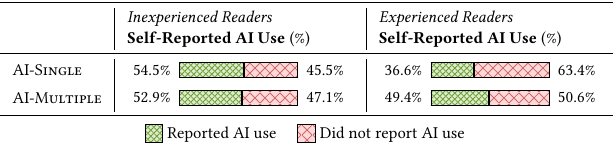}
\caption{Percentage of self-reported AI use among Inexperienced and Experienced Readers in the \singleai and \multipleai.}
\label{tab:ai_use_expertise_2}
\end{table*}


\subsection{How Did Participants Incorporate AI's Interpretation?}\label{copying_AI}

\subsubsection{Categorization of How Participants Incorporate AI Interpretation}

To understand how participants might incorporate AI interpretation(s), we examined textual overlap between participant answer and the AI interpretations they were exposed to, as well as copy-paste behavior. \revise{Textual overlap rate measures lexical similarity: we split both the participant answer and the AI interpretations into sentences, then broke each sentence into individual words. For each participant sentence, we identified the AI sentence with the greatest word overlap and counted the number of shared words. We summed these counts across all participant sentences and divided by the total word count of the participant's answer to obtain a rate. Copy rate captures explicit copying: we logged copy-paste events during the task, verified that copied fragments originated from the AI interpretations, identified matching words in the participant answer, and divided by the total word count.}
We primarily relied on textual overlap rate; as copying behavior can be incidental, we used copy rate as an additional metric for understanding explicit copying.
We identified three different categories: 
(1) complete overlap (instances where participants' interpretations were identical to the AI interpretation),
(2) high overlap (instances where textual overlap exceeded the medium rate of 57.1\% among all participants in the \singleai and \multipleai), and 
(3) low overlap (instances below the medium textual overlap rate). 

\definecolor{overlap}{RGB}{255,255,0}
\newcommand{\highlightcopy}[1]{\colorbox{overlap!20}{#1}}

\definecolor{overlap}{RGB}{255,255,0}
\newcommand{\highlightoverlap}[1]{\colorbox{overlap!40}{#1}}

\vspace{1em}
\noindent\fcolorbox{gray!70}{gray!10}{%
\begin{minipage}{1\linewidth}
\textbf{AI interpretation} \hfill \\
- Score: Feature (1), Interpretation (4), Writing (4) \\
\textbf{[Stylistic Feature]} \textit{The word/phrase that stuck out to me was ``dangerous banks''.} \textbf{[Explanation]} \textit{The phrase ``dangerous banks'' is used by the poet to emphasize the precariousness and risk associated with the edge of the creek, which can be seen as a metaphor for the boundaries within a relationship. By choosing these words, the poet underscores the tension and unpredictability inherent in love, suggesting that the reader should be cautious but also aware of the thrilling yet uncertain nature of deep emotional connections.}
\end{minipage}%
}
\vspace{1em}

In these instances of complete overlap, participants extensively copied text from AI, with 86.1\% having direct copy-and-paste behavior. They typically reproduced the AI interpretation verbatim. For example, P233 in the \singleai copied the entire AI interpretation, yielding the same good Performance score as the AI. 

Among high overlap instances, the majority of participants did not copy any text from the AI interpretation (59.3\%), but they heavily followed the AI.
For example, P20 in \singleai, without direct copy-and-pasting, lifted the core ideas from the AI with some rewording of the sentence structure. In the following example, the textual overlap is highlighted. 

\colorlet{mylightyellow}{yellow!20}  
\sethlcolor{mylightyellow}

\vspace{1em}
\noindent\textbf{Example of High Overlap} (P20, \singleai) \hfill \\ \text{- Score: Feature (1), Interpretation (4), Writing (5)}\\
\textbf{[Stylistic Feature]} \textit{\hl{The word/phrase that stuck out to me was ``dangerous banks''.}} \textbf{[Explanation]} \textit{The poet used the term to convey that there is \hl{risk associated with the edge of the creek}, \hl{which can be seen as a metaphor} that relates to the broader message of \hl{the boundaries within a relationship}. By using the phrase in this way, the poet tried to convey \hl{the tension and unpredictability inherent in love} to the readers that they \hl{should be cautious but also aware of the thrilling yet uncertain nature of deep emotional connections}.}

\vspace{1em}

In low overlap instances, most participants wrote their own interpretations without any copying (88.9\%). When copying or textual overlap occurred, they typically involved only brief stylistic examples or format elements. For example, P178 in \singleai only copied parts of the response format from the AI interpretation, and explained a different stylistic feature. In the following example, the copied text is highlighted. 

\colorlet{mylightred}{red!15} 
\sethlcolor{mylightred}

\vspace{1em}
\noindent\textbf{Example of Low Overlap} (P178, \singleai) \hfill \\
\text{- Score: Feature (1), Interpretation (2), Writing (2)}\\
\textbf{[Stylistic Feature]} \textit{\hl{The word/phrase that stuck out to me was} ``every scruple''.} \textbf{[Explanation]} \textit{This catches my attention because it shows how the water and by comparison, love, sweeps away not just objects but also doubts or careful thinking. It makes love feel wild and unstoppable, like it ignores rules.}
\vspace{1em}

These observations reveal a spectrum of AI use strategies, from wholesale adoption to selective borrowing to independent creation, each with distinct implications for both performance outcomes and subjective experience.

\subsubsection{Summary of Key Measures by How Participants Incorporate AI's Interpretation}

\begin{table*}[!htbp]
\centering
\includegraphics[width=0.9\textwidth]{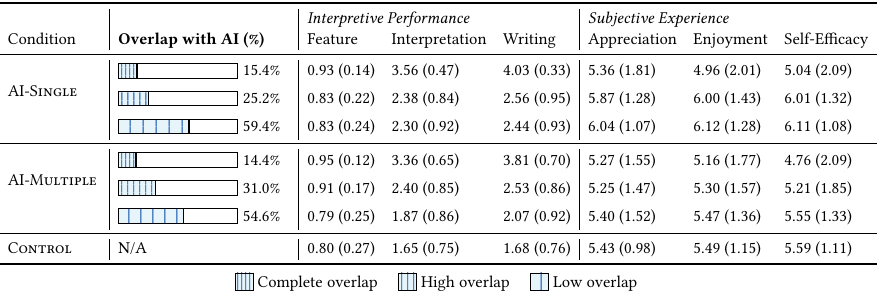}
\caption{Percentage of instances of different extent of textual overlap (complete overlap, high overlap, and low overlap), and summaries of Interpretive Performance and Subjective Experience measures (means with standard deviations in brackets). The \noai is included for reference.}
\label{tab:overlap_categories}
\end{table*}

Table \ref{tab:overlap_categories} reveals a consistent trade-off between performance and experience. Participants with high textual overlap achieved substantially better Interpretive Performance scores, yet had lower Subjective Experience ratings. This pattern was most pronounced in Complete overlap instances, where these participants achieved the highest Interpretive Performance scores, but the lowest Subjective Experience ratings, particularly in the \multipleai. Conversely, participants with Low overlap had the highest Subjective Experience ratings despite lower Interpretive Performance scores. 
This inverse relationship between performance and experience suggests that closely following AI interpretations improves objective performance on the surface, but at the expense of personal satisfaction and confidence. 


%% file: 6.discussion.tex
\section{Discussion}


\subsection{Key Findings}\label{interpretation_key_takeaways}

\subsubsection{AI Assistance Enhanced Interpretive Performance for Lay Readers.}
AI assistance either by presenting single or multiple interpretations significantly improved participants' interpretive performance.
However, even with improvements, participants' average scores in the \singleai (Feature: 0.85, Interpretation: 2.52, Writing: 2.71) and in the \multipleai (Feature: 0.85, Interpretation: 2.25, Writing: 2.46) remained below the AI's performance scores (graded by our graders using the same rubrics applied to the participants; Feature: 1.00, Interpretation: 4.00, Writing: 4.07).
Only participants who had complete textual overlap with AI interpretations approached these AI benchmarks in interpretation and writing quality (Table \ref{tab:overlap_categories}).


This performance gap suggests that for lay readers interpreting poetry, their baseline abilities are substantially lower than what AI can achieve through task automation. \citet{inkpen2023advancing} demonstrated that users whose baseline performance approximates AI capabilities show the greatest variability in AI assistance effectiveness. In contrast, our lay readers, whose baseline performance was well below AI, consistently benefited from AI assistance in terms of interpretive performance rather than variable effects. This indicates that AI assistance may be beneficial for close reading performance when there is a clear competency gap between human readers and AI capabilities.

\subsubsection{Exposure to Single AI Interpretation Enhanced Both Performance and Pleasure}

Exposure to a single AI interpretation improved both how well participants interpreted poems and the pleasure they derived from this task, as compared with the \noai condition. \revise{Particularly, this benefit still accrued to participants who reported not using the AI assistance.} Half of those in the \singleai condition reported not using it, yet they still performed better than the \noai group while maintaining the highest subjective experience ratings of any group in our study (Section \ref{self_reported_AI_use}). 
This pattern suggests that performance gains may be driven by ambient influence rather than direct adoption of the AI interpretation. Exposure to a sample interpretation may establish a quality benchmark that readers internalize, even when they consciously choose to pursue their own interpretation and not to use the AI. This aligns with the concept of ``abstract modeling,'' where observers extract underlying principles from a model through implicit learning without conscious awareness, rather than copying specific behaviors \cite{bandura1986social, zimmerman1997developmental, reber1993implicit, kihlstrom1987cognitive}. 
\revise{Participants may have chosen not to use AI to preserve autonomy and authenticity. For example, P24 in the \singleai noted in their open-ended responses reflecting their AI use: \textit{``my responses are authentically my own.''} P71 in the \singleai, while finding AI useful, emphasized that it served \textit{``as a means to further understanding; not do the work for me,''} suggesting that readers can benefit from AI exposure while still maintaining a sense of independent engagement.}

This suggests that, for lay cultural interpretation, modest AI assistance (e.g., presenting a single AI interpretation) can provide an optimal level of support, which is enough to elevate their interpretive reasoning while increasing their pleasure from appreciation, enjoyment, and self-efficacy in the process. This benefit means that such modest AI assistance can help readers when they feel stuck, while still leaving room to discover personal meaning, resonate with the text on their own terms, and feel competent in the process. As some described, the AI was \textit{``helpful as a prompt to go deeper''} (P2, \singleai) and useful \textit{``to cross-reference my own interpretations''} (P10, \singleai), providing a reference point rather than displacing the meaning-discovery process. This type of support can enhance rather than undermine participants' sense of competence, as \textit{``a collaborative partner in the close reading process to articulate my thoughts and to increase confidence''} (P32, \singleai).

\subsubsection{Exposure to Multiple AI Interpretations Enhanced Performance, But Not Pleasure}

AI assistance by presenting multiple AI interpretations improved Interpretive Performance but didn't enhance any aspect of Subjective Experience. For relatively experienced readers, it even reduced their Self-Efficacy. This suggests that more AI assistance does not translate to greater pleasure---a divergence that is crucial for understanding AI's role in lay cultural interpretation, where the primary goal for most readers is not to produce \textit{the} optimal analysis, but to meaningfully engage with a text for one's personal discovery.

\revise{Several mechanisms may explain this result.} First, multiple AI interpretations may have threatened rather than supported participants' sense of competence. \revise{As participants noted in their open-ended responses reflecting on
their AI use, some felt they could not match the AI's sophistication: \textit{``AI's understanding of these poems was much deeper and, quite frankly, better than mine. I just couldn't compete''} (P26, \multipleai). Others reported that seeing multiple AI interpretations undermined their confidence: \textit{``It made me feel less confident about how I was interpreting the poems''} (P126, \multipleai). Some experienced AI as a competitor rather than a resource: \textit{``a threat that I had to do better than''} (P35, \multipleai).}
Second, multiple AI interpretations may have exhausted the interpretive space, diminishing the intrinsic reward of personal discovery. \revise{When AI interpretations cover multiple angles, some felt there was nothing left for them to contribute after seeing the AI's comprehensive coverage: \textit{``It had answered so in depth that there was not anything else for me to say on the matter''} (P24, \multipleai).}


Behavioral patterns further support these potential mechanisms. Viewing behavior (Section \ref{viewing_AI_multiple}) reveals that less than half of participants looked at all three available AI interpretations, with many viewing only one. This limited engagement may reflect an autonomy preservation strategy---avoiding AI content to protect space for personal meaning-making. Additionally, participants in the \multipleai condition spent more time in the AI area on average and less time engaging with the poem itself (Section \ref{time_allocation}), suggesting they shifted focus from the poem to the AI interpretations.

\subsubsection{The Performance-Pleasure Trade-off Among Heavy Copiers}

\revise{Beyond our main analyses, behavioral engagement patterns revealed a key tension:} participants who reported using AI or whose responses showed high textual overlap with AI interpretations achieved performance scores closer to the AI benchmark but consistently reported lower Subjective Experience ratings across Enjoyment, Appreciation, and Self-Efficacy measures. Participants with complete textual overlap achieved average Interpretation and Writing scores near the AI benchmark of 4.0, but their pleasure ratings were the lowest across all groups. This performance-pleasure trade-off among heavy copiers suggests that while heavy reliance offers one pathway to higher performance, it may come at the cost of the intrinsic rewards that make close reading meaningful.


\revise{The gradual release of responsibility model \cite{pearson1983instruction, fisher2021better, duke2009effective, bandura1986social, zimmerman1997developmental} offers a lens for understanding this trade-off. This model describes effective learning as a progression from observing a teacher's demonstration to practicing independently. In our study, AI assistance may have functioned like a teacher's demonstration: the heavy copiers may have remained in the observation phase, never making this transition. Yet participants with low textual overlap or who report not using AI might use AI as a teacher's demonstration while transitioning to independent discovery of textual meaning and thus allowing for the emergence of personal satisfaction.}
\revise{Participants' open-ended responses reflecting on their AI use can support this interpretation. Participants might have heavily relied on AI due to compensatory or pressured motivations, using AI not as a collaborative tool but because exposure to AI interpretation undermined their confidence or left them feeling they had nothing original to contribute. Some turned to AI because it \textit{``made me feel less confident about how I was interpreting the poems''} (P126, \multipleai). Some felt AI had already exhausted the interpretive space: \textit{``It had answered so in depth that there was not anything else for me to say''} (P24, \multipleai).} When AI displaces rather than supports personal meaning-making, the surface goal of a high-quality interpretation may be achieved, but the internal rewards of discovery are lost. \revise{In contrast, those who chose not to use AI articulated reasons tied to preserving autonomy and protecting the joy of discovery. Some chose to avoid AI because they wanted to \textit{``connect with the poems on a personal level''} (P109, \singleai). Others felt that using AI would compromise their authenticity: \textit{``I would've felt like a fraud not using my own thinking''} (P81, \singleai).}

\subsection{Implications for Human-AI Interaction in Culture and Arts}\label{design_implications}



\subsubsection{Design Implications}

Our findings offer design implications for AI-assisted cultural interpretation that support both performance and pleasure.
The reward of lay cultural consumption is fundamentally different from academic or professional contexts. While AI can help produce more sophisticated interpretations, this performance gain becomes less meaningful if it diminishes the pleasure, discovery, and personal meaning-making that motivate recreational reading. As AI becomes increasingly capable of reasoning, there's a risk of outsourcing not just the cognitive work but the experiential value of interpretation. The performance-pleasure trade-off we observed suggests that the automation of cultural interpretation alone may miss what makes it valuable.

\revise{Based on our findings, we suggest an overarching design principle:} \textit{``less is more.''} Designers should resist presenting maximal AI capabilities and instead carefully calibrate support to maintain human engagement, even if this means accepting lower objective performance. 
\revise{For cultural interpretation, AI assistance should leave sufficient space for personal discovery and avoid presenting overly comprehensive coverage. Exhaustive AI analysis may leave readers feeling they have nothing original to contribute. Research on choice overload demonstratea that extensive options can reduce satisfaction and motivation rather than enhance them \cite{iyengar2000choice}. Notably, this concern applies not only to the number of AI interpretations but also to their coverage and depth: a single thorough interpretation that covers multiple angles, identifies all key features, and explains their effects comprehensively can be just as overwhelming as multiple separate interpretations. Cognitive load theory suggests that extraneous information can reduce cognitive resources available for meaningful engagement \cite{sweller1988cognitive}. Therefore, rather than providing complete analysis, AI assistance might better serve readers by offering a starting point or partial insight, enough to scaffold understanding without foreclosing the reader's own interpretive journey.}

\looseness=-1 \revise{Moreover, designers could make AI assistance available after an initial period of independent engagement. Our findings suggest that readers can naturally regulate their use of AI assistance to preserve the pleasure of close reading. Recent work by \citet{qin2025timing} found that providing delayed AI assistance preserved the originality of creative ideas compared to having AI assistance throughout the task. Self-determination theory suggests that preserving autonomy is essential for intrinsic reward \cite{ryan2000self}. In the context of cultural interpretation, delaying AI assistance may allow readers to experience the autonomous engagement that makes close reading pleasurable. Design can encourage readers to attempt their own close reading before turning to AI assistance. For example, a poetry platform could pop up AI assistance only after a reader has spent some time looking at a poem. This delay can allow readers to benefit from AI scaffolding while still experiencing the satisfaction of independent discovery.}



\subsubsection{Broader Implications}


These implications extend beyond poetry to other forms of cultural work. While our study focused on poems as text-based cultural works, similar dynamics may apply to other media, such as visual (paintings), auditory (music, songs), or multimodal (films, theater, video games). In each case, AI integration should also consider not only whether AI helps people understand culture better, but also whether it preserves the human experience of discovering meaning for themselves. The success of \textit{``less is more''} for AI assistance of lay readers close reading of poetry suggests that thoughtfully calibrated AI support may better serve both human capabilities and human experience.

\subsection{Limitations and Future Work}\label{limitations}



\subsubsection{Forms of AI Assistance}
In our study, we provided AI assistance in the form of pre-generated interpretations rather than a freeform AI chatbot. This decision involved important trade-offs. We chose pre-generated interpretations to control that participants saw interpretations of similar quality that we could validate beforehand. This allowed us to isolate the effects of the number of AI interpretations from variations in AI response quality or individual participants' prompting abilities. This approach does not reflect how AI assistance for cultural interpretation might evolve. 
Given that our research aims to advance discussion on whether and how AI should be used to assist cultural interpretation, we prioritized understanding the effects of AI exposure over the mechanics of AI interaction. Future work could explore how people engage with conversational AI for cultural interpretation. 

\revise{In addition, our interface for the \multipleai condition required participants to click to reveal additional interpretations. Future work could compare different interface design choices (e.g., showing multiple interpretations altogether) to understand how they might affect participants' performance and subjective experience.}

\subsubsection{Poems Selected}

Our study examined three poems selected for their accessibility and thematic diversity. While these poems represented different styles and themes (love, nature, social identity), they may not capture the full range of poetic forms, cultural contexts, or difficulty levels that readers encounter. Future studies could test an even broader range of poems to understand how AI assistance affects close reading across diverse literary styles.

\subsubsection{\revise{Categorization of Inexperienced and Experienced Readers}}

Our binary classification of participants as inexperienced and experienced readers based on humanities coursework provides one indicator of close reading expertise. Future research could develop more nuanced categorization schemes that capture multiple dimensions of expertise.


\subsubsection{A Focus on Lay Readers and Everyday Contexts}
Our study deliberately focused on lay readers engaging with poetry for personal enrichment rather than students in educational settings or professionals conducting literary analysis. This population represents the vast majority of people encountering cultural content online, yet their needs and goals differ substantially from educational contexts. Our finding that modest AI assistance benefits both performance and pleasure while extensive assistance undermines experiential value may not generalize beyond this recreational context.

\subsubsection{\revise{Operationalization of Pleasure.}}
\revise{We operationalized the pleasure of close reading through three subjective experience constructs---appreciation, enjoyment, and self-efficacy---based on close reading literature \cite{abrams1999glossary,guillory2025close,bialostosky2006should} and intrinsic reward theory \cite{csikszentmihalyi2015intrinsic,csikszentmihalyi2014play}. 
However, individuals may experience or define pleasure in close reading differently than what the broad literature suggests. Future work could explore what dimensions readers perceive as constituting pleasure in close reading.}

\revise{We used single-item measures for each construct to reduce participant fatigue given the study length. While common in large-scale studies with repeated measures, single-item measures capture less nuanced variation than multi-item scales \cite{schwarzer1995generalized}. Future work examining specific facets of subjective experience should adapt such scales for more precise measurement.}

\subsubsection{Other Directions for Future Work}

\looseness=-1 We intentionally did not include a practice-and-test design where AI assistance would be removed to measure learning gains. While such designs are valuable for educational research, they don't reflect how lay readers engage with cultural texts online, where learning gains are not the main objective. Future research focused on educational contexts could investigate whether exposure to AI interpretations leads to lasting improvements in interpretive skills when AI is subsequently removed.

\revise{Our study did not provide performance-based incentives, which might encourage deeper engagement with both the poems and AI interpretations. Future work could explore how incentive designs can influence participants' performance and engagement in cultural interpretation with AI assistance.}

\revise{Our study focused on lay readers and the foundational observation step of close reading---identifying stylistic features and explaining their effects within the text---based on the CRIT framework \cite{utaustin2024critical}. This focus enabled consistent measurement as it does not heavily depend on external context and background knowledge. Broader close reading practices can involve connecting texts to historical, biographical, or theoretical contexts. Future work could examine how AI assistance affects these more advanced interpretive stages with students in professionals or educational settings.}


%% file: 7.conclusion.tex
\section{Conclusion}

We examined how AI assistance in the forms of single and multiple AI interpretations affect close reading poems in a controlled experiment with 400 participants. While both single and multiple AI interpretations improved interpretative performance, their effects on subjective experience diverged sharply: single AI interpretation enhanced appreciation, enjoyment, and self-efficacy, whereas multiple interpretations provided no subjective benefits and even decreased self-efficacy among experts. These findings reveal a fundamental insight for AI-assisted cultural interpretation: more exposure to AI assistance does not necessarily create better experiences. As AI becomes increasingly integrated into cultural engagement online, our results suggest that the role of AI in cultural interpretation requires careful calibration between performance enhancement and experiential value. We call for continued investigation into how AI can support rather than supplant the human experience of discovering meaning in cultural works.

%% file: 8.appendix.tex
\section{Appendix}

\begin{table*}[h]
\centering
\small
\begin{tabular}{llllllll}
\toprule
Group & Measure & Condition & Mean (SD) & Cohen's d & p-value & Sig. \\
\midrule
\multirow{9}{*}{Inexperienced Readers} 
& \multirow{3}{*}{Feature} 
& \noai & 0.779 (0.230) & -- & -- & -- \\
& & \singleai & 0.822 (0.169) & 0.208 & 0.529 & ns \\
& & \multipleai & 0.824 (0.190) & 0.212 & 0.277 & ns \\
\cmidrule{2-7}
& \multirow{3}{*}{Interpretation} 
& \noai & 1.617 (0.810) & -- & -- & -- \\
& & \singleai & 2.564 (0.912) & 1.106 & 0.000 & *** \\
& & \multipleai & 2.153 (0.891) & 0.630 & 0.000 & *** \\
\cmidrule{2-7}
& \multirow{3}{*}{Writing} 
& \noai & 1.656 (0.784) & -- & -- & -- \\
& & \singleai & 2.749 (1.013) & 1.229 & 0.000 & *** \\
& & \multipleai & 2.370 (1.014) & 0.791 & 0.000 & *** \\
\midrule
\multirow{9}{*}{Experienced Readers} 
& \multirow{3}{*}{Feature}
& \noai & 0.818 (0.189) & -- & -- & -- \\
& & \singleai & 0.870 (0.171) & 0.289 & 0.101 & ns \\
& & \multipleai & 0.896 (0.155) & 0.337 & 0.059 & ns \\
\cmidrule{2-7}
& \multirow{3}{*}{Interpretation} 
& \noai & 1.684 (0.553) & -- & -- & -- \\
& & \singleai & 2.474 (0.797) & 1.166 & 0.000 & *** \\
& & \multipleai & 2.328 (0.867) & 0.874 & 0.000 & *** \\
\cmidrule{2-7}
& \multirow{3}{*}{Writing} 
& \noai & 1.696 (0.604) & -- & -- & -- \\
& & \singleai & 2.678 (0.900) & 1.299 & 0.000 & *** \\
& & \multipleai & 2.541 (0.936) & 1.061 & 0.000 & *** \\
\bottomrule
\end{tabular}
\caption{Interpretive Performance by expertise group and condition. Means and standard deviations are descriptive statistics from participant-level averages. Cohen's d and p-values are from Mann-Whitney U tests comparing \noai vs. AI-Single and \noai vs. \multipleai within each expertise group. P-values are Holm-Bonferroni adjusted. Significance: *** $p < 0.001$, ** $p < 0.01$, * $p < 0.05$, ns = not significant; -- = reference group (Control).}
\label{tab:expertise_group_performance}
\end{table*}

\begin{table*}[h]
\centering
\small
\begin{tabular}{llllllll}
\toprule
Group & Measure & Condition & Mean (SD) & U & r & p-value & Sig. \\
\midrule
\multirow{9}{*}{Inexperienced Readers} 
& \multirow{3}{*}{Appreciation} 
& \noai & 5.319 (1.374) & -- & -- & -- & -- \\
& & \singleai & 6.164 (0.795) & 1179.5 & 0.385 & 0.001 & ** \\
& & \multipleai & 5.571 (1.162) & 2142.5 & 0.086 & 0.978 & ns \\
\cmidrule{2-8}
& \multirow{3}{*}{Enjoyment} 
& \noai & 5.356 (1.147) & -- & -- & -- & -- \\
& & \singleai & 5.964 (0.888) & 1356.0 & 0.315 & 0.009 & ** \\
& & \multipleai & 5.328 (1.117) & 2440.0 & 0.012 & 0.907 & ns \\
\cmidrule{2-8}
& \multirow{3}{*}{Self-efficacy} 
& \noai & 5.338 (1.228) & -- & -- & -- & -- \\
& & \singleai & 5.921 (1.087) & 1343.0 & 0.322 & 0.007 & ** \\
& & \multipleai & 5.353 (1.256) & 2443.0 & 0.013 & 0.897 & ns \\
\midrule
\multirow{9}{*}{Experienced Readers} 
& \multirow{3}{*}{Appreciation}
& \noai & 5.708 (1.090) & -- & -- & -- & -- \\
& & \singleai & 5.839 (1.099) & 1752.5 & 0.101 & 0.978 & ns \\
& & \multipleai & 5.379 (1.422) & 2691.5 & 0.090 & 0.978 & ns \\
\cmidrule{2-8}
& \multirow{3}{*}{Enjoyment} 
& \noai & 5.633 (1.149) & -- & -- & -- & -- \\
& & \singleai & 5.861 (1.114) & 1762.0 & 0.149 & 0.287 & ns \\
& & \multipleai & 5.407 (0.997) & 3103.5 & 0.168 & 0.235 & ns \\
\cmidrule{2-8}
& \multirow{3}{*}{Self-efficacy} 
& \noai & 5.845 (0.917) & -- & -- & -- & -- \\
& & \singleai & 5.917 (1.141) & 1752.0 & 0.154 & 0.259 & ns \\
& & \multipleai & 5.307 (1.308) & 3338.0 & -0.257 & 0.021 & * \\
\bottomrule
\end{tabular}
\caption{Subjective Experience by expertise group and condition. Means and standard deviations are descriptive statistics from participant-level averages. Cohen's d and p-values are from Mann-Whitney U tests comparing \noai vs. AI-Single and \noai vs. \multipleai within each expertise group. P-values are Holm-Bonferroni adjusted. Significance: *** $p < 0.001$, ** $p < 0.01$, * $p < 0.05$, ns = not significant; -- = reference group (Control).}
\label{tab:subjective_measures}
\end{table*}